\documentclass[pra,twocolumn,10pt,aps,showpacs,amsmath,amsfonts,floatfix,letterpaper,superscriptaddress]{revtex4-1}

\usepackage{times}
\usepackage{txfonts}

\usepackage{amsmath}
\usepackage{amssymb}
\usepackage{hyperref}
\usepackage{multirow}

\newcommand{\ee}{\mathrm{e}}
\newcommand{\ii}{\mathrm{i}}

\newcommand{\del}{\vec{\nabla}}
\newcommand{\Av}{\vec{A}}
\newcommand{\av}{\vec{a}}

\newcommand{\deltav}{\vec{\delta}}
\newcommand{\zerom}{\boldsymbol{0}}

\newcommand{\vphiv}{\boldsymbol{\varphi}}

\newcommand{\rv}{\vec{r}}
\newcommand{\Rv}{\vec{R}}
\newcommand{\mv}{\vec{m}}
\newcommand{\Tv}{\vec{T}}
\newcommand{\Sv}{\vec{S}}
\newcommand{\Nv}{\vec{N}}
\newcommand{\Bv}{\vec{B}}
\newcommand{\zerov}{\vec{0}}
\newcommand{\rhov}{\vec{\rho}}

\newcommand{\bigO}{\mathcal{O}}

\newcommand{\zm}{\boldsymbol{z}}

\newcommand{\sigmamv}{\vec{\boldsymbol{\sigma}}}

\newcommand{\scG}{\mathcal{G}}
\newcommand{\sigmam}{\boldsymbol{\sigma}}

\newcommand{\scH}{\mathcal{H}}

\newcommand{\scS}{\mathcal{S}}
\newcommand{\scZ}{\mathcal{Z}}

\newcommand{\scT}{\mathcal{T}}
\newcommand{\scC}{\mathcal{C}}

\newcommand{\beq}[1]{\begin{equation}\label{#1}}
\newcommand{\eeq}{\end{equation}}
\newcommand{\refeq}[1]{Eq.~(\ref{#1})}

\newcommand{\refeqand}[2]{Eqs.~(\ref{#1}) and (\ref{#2})}
\newcommand{\refcite}[1]{Ref.~\onlinecite{#1}}

\newcommand{\reffig}[1]{Fig.~\ref{#1}}

\newcommand{\refsec}[1]{Section~\ref{#1}}
\newcommand{\refsecand}[2]{Sections~\ref{#1} and \ref{#2}}
\newcommand{\refsecs}[2]{Sections~\ref{#1}--\ref{#2}}

\newcommand{\punc}[1]{\,{\text{#1}}}
\newcommand{\sub}[1]{_{\text{#1}}}

\newcommand{\blp}{\boldsymbol{(}}
\newcommand{\brp}{\boldsymbol{)}}

\newcommand{\SU}{\mathrm{SU}}
\newcommand{\U}{\mathrm{U}}

\newcommand{\NCP}{NC$CP^1$}
\newcommand{\JQ}{\(JQ\)}

\DeclareMathOperator{\Div}{div}

\newcommand{\hsmash}[1]{\!\!\!\!\!\!\!#1\!\!\!\!\!\!\!}

\usepackage{color}

\usepackage{graphicx}
\usepackage{ifpdf}
\ifpdf

\else

\fi

\begin{document}

\title{Scaling dimensions of higher-charge monopoles at deconfined critical points}

\author{G. J. Sreejith}
\affiliation{Max Planck Institute for Physics of Complex Systems, Dresden, Germany}

\author{Stephen Powell}
\affiliation{School of Physics and Astronomy, The University of Nottingham, Nottingham, NG7 2RD, United Kingdom}

\begin{abstract}

The classical cubic dimer model has a columnar ordering transition that is continuous and described by a critical Anderson--Higgs theory containing an $\SU(2)$-symmetric complex field minimally coupled to a noncompact $\U(1)$ gauge theory. Defects in the dimer constraints correspond to monopoles of the gauge theory, with charge determined by the deviation from unity of the dimer occupancy. By introducing such defects into Monte Carlo simulations of the dimer model at its critical point, we determine the scaling dimensions \(y_2 = 1.48\pm0.07\) and \(y_3 = 0.20\pm0.03\) for the operators corresponding to defects of charge \(q=2\) and \(3\) respectively. These results, which constitute the first direct determination of the scaling dimensions, shed light on the deconfined critical point of spin-$\frac{1}{2}$ quantum antiferromagnets, thought to belong to the same universality class. In particular, the positive value of \(y_3\) implies that the transition in the \JQ\ model on the honeycomb lattice is of first order.

\end{abstract}

\maketitle

\section{Introduction}

One of the most remarkable consequences of the theory of critical phenomena is universality, the occurrence of identical values for nontrivial measurable quantities at phase transitions in quite different physical systems \cite{LandauBook,FisherReview}. For Landau transitions, whose critical properties are described by the long-wavelength fluctuations of an order parameter, two transitions usually belong in the same universality class when their space(--time) dimensionalities are the same, and when their order parameters have identical symmetry properties at the critical point.

In certain unconventional phase transitions, the order parameter is not the basic field describing the critical properties, but can instead be expressed as a compound object in terms of the basic fields \cite{Senthil}. Universality can be particularly striking in such cases, uniting transitions where the microscopic models and phenomenology bear little resemblance.

The focus of the current work is the noncompact $CP^{1}$ (\NCP) universality class, proposed as describing a number of unconventional phase transitions. These include: the N\'eel--valence-bond solid (VBS) transition \cite{Senthil} in the \JQ\ model \cite{SandvikJQ} (a quantum antiferromagnet with frustration favoring singlet dimers) on certain two-dimensional (2D) lattices; an ordering transition in the 3D Heisenberg model with suppression of ``hedgehog'' defects \cite{Motrunich}; the loop-proliferation transition in certain 3D classical loop models \cite{Nahum1,Nahum2}; and a columnar ordering transition in the classical cubic dimer model (CDM) \cite{Alet,CubicDimersPRL,Charrier,Chen,CubicDimersPRB,Charrier2,Sreejith}. The \NCP{} theory involves a noncompact $\U(1)$ gauge theory minimally coupled to an $\SU(2)$-symmetric complex field and driven through an Anderson--Higgs transition; order parameters for the transitions can be constructed as combinations of these fields. While considerable debate remains about the true nature of the phase transitions, and many aspects are not yet understood satisfactorily \cite{Troels2013,Nahum2}, it seems clear that there exists at least a large range of length scales over which their properties are well described by the \NCP{} model.

Based on this viewpoint, we exploit universality to provide results for quantities of central importance for the transition in the \JQ\ model, using numerical simulations performed on the CDM. The quantities in question are the scaling dimensions, or equivalently renormalization-group (RG) eigenvalues \(y_q\), of operators that insert magnetic monopoles of charge $q$. Monopoles are absent in the \NCP{} theory---according to an argument due to Polyakov \cite{Polyakov}, they always occur at nonzero density in a \emph{compact} $\U(1)$ gauge field but are absent when the field is noncompact---but provide a diagnostic for the transition: An oppositely charged pair of test monopoles is deconfined in the Coulomb (disordered) phase, but becomes confined in the Higgs (ordered) phase.

In the CDM, monopoles are simply defects in the dimer constraint, at which dimers overlap or a site is unoccupied, and which carry charge in an effective gauge-theoretical description  \cite{Huse,HenleyReview}. In the \JQ\ model, by contrast, they correspond to topologically nontrivial hedgehog configurations (in 3D space--time) of the degrees of freedom. While such defects generically occur at nonzero density, they are in some cases irrelevant at the transition, which can therefore be described by a noncompact (monopole-free) critical theory. This comes about because of quantum Berry phases that are associated with hedgehogs and endow monopoles with nontrivial transformation properties under the lattice symmetries. (They can therefore be associated with VBS order; see \refcite{Senthil} and \refsec{SectionDQC}.) Monopoles of charge $q$ are therefore suppressed at the (symmetric) critical point, unless $q$ is an integer multiple of $q\sub{min}$, determined by the rotation symmetry of the lattice. In order for the transition to be described by the \NCP{} theory, in which all monopoles are forbidden, these remaining monopoles must be irrelevant (i.e., $y_q < 0$ for all nonzero $q \in q\sub{min}\mathbb{Z}$) at the appropriate RG fixed point.

It is therefore desirable to have a method of finding the scaling dimensions of charge-$q$ monopoles at the \NCP{} critical point. Due to the topological character of hedgehogs, these quantities are difficult to extract directly in quantum spin models. So far, only indirect evidence for their relevance or irrelevance has been found by determining the order of the transition in the \JQ\ model on lattices of different symmetries \cite{Block}, and by studying the scaling of powers of the VBS order parameter \cite{Harada}. The scaling dimensions can also in principle be calculated by generalizing the spin model to $\SU(N)$ symmetry and performing a large-$N$ expansion, but the calculations are technically demanding and only results to low order in $N^{-1}$ are available \cite{MurthySachdev,Dyer}.

Because monopoles have a simple representation in the classical CDM, and because the latter model is particularly amenable to Monte Carlo (MC) simulations, it provides an efficient route to calculation of the quantities $y_q$ for the \NCP{} universality class. The goal of this work is to demonstrate that $y_q$ can be found using such simulations and to give explicit values, amounting to the first direct numerical calculation of these exponents.

\subsection*{Outline}

In \refsec{SecCDM}, we introduce the cubic dimer model and describe its phase structure. In \refsec{SectionDQC}, we review the theory of the N\'eel--VBS transition in quantum spin models, emphasizing the effect of monopoles on the critical properties. \refsec{SecMethods} describes the scaling of the monopole distribution function in the dimer model, and how this can be determined in numerical studies. We then describe, in \refsec{SecNumericalMethods}, the MC algorithms that we use to study the behavior of monopoles of charge-\(q\) in the dimer model. We present our numerical results in \refsec{SecResults} before concluding, with discussion of the implications for the critical properties of quantum spin models, in \refsec{SecConclusions}.

\section{Cubic dimer model}
\label{SecCDM}

In this section, we briefly review the definition, phase structure, and critical properties of the CDM.

\subsection{Definition}

Our numerical studies treat a classical statistical model of hard-core dimers on the links of a cubic lattice. The dimer occupation number $d_\mu(\rv) \in \{0,1\}$ is defined as the number of dimers on the link joining the site $\rv$ to its neighbor $\rv+\deltav_\mu$ in the direction $\mu$ (where \(\deltav_\mu\) are the unit vectors in the 3 directions). Apart from the defect sites (see \refsec{SecCDMDefects}), the number of dimers touching site $\rv$,
\beq{EqDefinen}
n(\rv) = \sum_\mu [d_\mu(\rv) + d_\mu(\rv - \deltav_\mu)]\punc{,}
\eeq
is constrained by $n(\rv) = 1$.

To study the transition to an ordered state, we introduce interactions that count the number of nearest-neighbor parallel dimers,
\beq{EqDefineN2}
N_2 = \sum_{\rv} \sum_{\substack{\mu \\ \nu \neq \mu}} d_\mu(\rv) d_\mu(\rv + \deltav_{\nu}) \punc{,}
\eeq
and the number of parallel dimers around cubes of the lattice,
\beq{EqDefineN4}
N_4 = \sum_{\rv} \sum_{\substack{\mu\\\nu \neq \mu\\\rho \neq \mu,\nu}} d_\mu(\rv) d_\mu(\rv + \deltav_{\nu})  d_\mu(\rv + \deltav_\rho) d_\mu(\rv + \deltav_{\nu} + \deltav_\rho)
\punc{;}
\eeq
the configuration energy is
\beq{EqDefineE}
E = v_2 N_2 + v_4 N_4\punc{.}
\eeq
The continuous transition of interest occurs for $v_2 < 0$ and $v_4 \ge 0$; in the following we choose units in which $v_2 = -1$. (The transition is also present for \(v_4 = 0\), but it is then found to show mean-field critical exponents, likely as the result of proximity to a multicritical point \cite{Charrier2}.) We use a lattice with periodic boundary conditions and an even number $L$ of sites in each direction.

\subsection{Phase structure and critical theory}
\label{CDMPhases}

At high temperature $T$, the dimer model exhibits a Coulomb phase, in which occupation-number correlations are algebraic and test monomers are deconfined. A continuum description for this phase can be found by first defining an effective magnetic field \cite{Huse,HenleyReview}
\beq{EqDefineB}
B_\mu(\rv) = \eta_{\rv}\left[d_\mu(\rv) - \frac{1}{6}\right]\punc{,}
\eeq
where $\eta_{\rv} = (-1)^{\sum_\mu r_\mu}$ is $\pm 1$ on the two sublattices. The lattice divergence, defined by
\beq{EqDefineDiv}
\Div_{\rv} B \equiv \sum_{\mu} [ B_\mu(\rv) - B_\mu(\rv - \deltav_\mu)]\punc{,}
\eeq
obeys
\beq{EqDivB}
\Div_{\rv} B = \eta_{\rv}[n(\rv) - 1]
\punc{,}
\eeq
and so vanishes in configurations obeying the dimer constraint. Defects in the constraint act as charges, or magnetic monopoles, under this discrete Gauss law, with sign depending on the sublattice.

The partition function \(\scZ\) for the dimer model, subject to the constraint, can be written in terms of \(B\) as
\beq{EqDefineZ}
\scZ = \sum_{\hsmash{\substack{\{B_\mu(\rv)\}\\ \Div_{\rv} B = 0}}} \ee^{-E/T}\punc{,}
\eeq
where the energy \(E\) of a configuration is expressed in terms of \(B\).

In a long-wavelength description, $B_\mu(\rv)$ is replaced by a continuum vector field $\Bv(\rv)$ with vanishing divergence, $\del\cdot\Bv = 0$. The effective action density in the Coulomb phase is \cite{Huse,HenleyReview}
\beq{EqCoulombPhaseAction}
\scS\sub{gauge} = \frac{1}{2}K\lvert \Bv \rvert^2 = \frac{1}{2}K\lvert \del\times\Av \rvert^2\punc{,}
\eeq
where $\Bv = \del\times\Av$, and higher-order corrections, redundant under the RG, have been omitted.

At a critical temperature $T\sub{c}(v_4)$ the system orders into a columnar dimer crystal, which breaks both lattice translation and rotation symmetries. A suitable order parameter is the ``magnetization'' \(\mv\), defined by
\beq{EqDefinem}
m_\mu = \frac{2}{L^3}\sum_{\rv} (-1)^{r_\mu} d_\mu(\rv)\punc{.}
\eeq
In this ordered phase, the (connected) correlations are short-ranged and test  monopoles are subject to confinement.

The simultaneous appearance of order and confinement can be described by adding vector matter fields $\vphiv$ to the gauge theory, so that the action density becomes
\beq{EqCriticalAction}
\begin{aligned}
\scS\sub{critical} &= \scS\sub{gauge} + \scS\sub{matter} + \scS\sub{matter--gauge}\\
\scS\sub{matter} &= s\lvert\vphiv\rvert^2 + u(\lvert\vphiv\rvert^2)^2\\
\scS\sub{matter--gauge} &= \lvert(\del - \ii \Av) \vphiv\rvert^2\punc{,}
\end{aligned}
\eeq
with pure gauge part given by \refeq{EqCoulombPhaseAction}. The matter field $\vphiv$ is minimally coupled to $\Av$ and so, being dual to the magnetic monopoles, can be viewed as carrying electric charge. The transition occurs when $s$ is tuned through its critical value $s\sub{c}$. For $s < s\sub{c}$, $\vphiv$ condenses and $\Av$ acquires an effective mass term by the Anderson--Higgs mechanism. This in turn eliminates the algebraic correlations and confines the monomers (magnetic charges) through the Meissner effect.

In the case of the dimer model, the matter field $\vphiv$ has $2$ components \cite{CubicDimersPRL,Charrier,Chen,CubicDimersPRB}, and the critical action is symmetric under $\SU(2)$. The vector structure of the order parameter is encoded in the $\SU(2)$ vector structure by \(\mv = \vphiv^\dagger \sigmamv \vphiv\), where \(\sigmam_\mu\) is a Pauli matrix. Replacing $\scS\sub{matter}$ by a fixed-length constraint on $\vphiv$, one arrives at the standard form for the \NCP{} theory.

\subsection{Defects at the critical point}
\label{SecCDMDefects}

A defect in the dimer model is a site \(\rv\) that is touched either by no dimers or by more than one, and so has \(n(\rv) \neq 1\). In terms of the effective magnetic field, such a defect has \(\Div_{\rv} B \neq 0\), and so can be viewed as a magnetic monopole. According to \refeq{EqDivB}, empty sites and those where two dimers intersect both have unit charge, with sign that alternates on the two sublattices. When defects occur at finite density, they destroy the topological order, and lead to a conventional phase transition between ordered and disordered phases \cite{Sreejith}.

By contrast, a test pair in an otherwise defect-free configuration can be used as a diagnostic of the phase transition, using the concept of confinement. This can be given a precise definition in the dimer model by considering first the partition function evaluated in the presence of a set of charges at fixed positions,
\beq{EqDefineZQ}
\scZ[Q_{\rv}] = \sum_{\hsmash{\substack{\{B_\mu(\rv)\}\\ \Div_{\rv} B = Q_{\rv}}}} \ee^{-E/T}\punc{.}
\eeq
The distribution function for a test pair of monopoles of charge \(\pm q\) at $\rv_{\pm}$ is then defined by the ratio of the partition function in their presence to that in their absence,
\beq{EqDefineGq}
G_q(\rv_+ - \rv_-) = \frac{\scZ[q\delta_{\rv,\rv_+} - q \delta_{\rv,\rv_-}]}{\scZ}\punc{.}
\eeq
This quantity is related to the effective interaction \(V_q^{\text{eff}}\) between the pair by \(G_q(\rv) = \exp \blp{-V_q^{\text{eff}}(\rv)/T}\brp\).

For \(T < T\sub{c}\), test charges are confined: \(G_q(\rv)\) decreases exponentially for large separation \(\rv\) (and any \(q\)), corresponding to a confining interaction, \(V_q^{\text{eff}} \propto \lvert \rv\rvert\). For \(T > T\sub{c}\), \(V_q^{\text{eff}}(\rv)\) has a finite limit, \(G_q(\rv)\) approaches a nonzero constant, and charges are deconfined.

The behavior of monopoles at the critical point can be understood by considering a real-space renormalization procedure that preserves the discrete nature of the charges. Following a similar logic to the case of unit-charge monopoles \cite{MonopoleScalingPRL,MonopoleScalingPRB}, one can identify a set of scaling dimensions \(y_q\) corresponding to monopoles of charge \(\pm q\). This implies that, at the critical point, the monopole distribution function takes the form
\beq{EqGqLargeR}
G_q(\Rv) \sim \lvert \Rv \rvert^{-2x_q}\punc{,}
\eeq
for large separation \(\Rv\), where \(x_q = d - y_q\).

It should be noted that multiple defects in close proximity are expected to act, with regard to long-wavelength properties, exactly as a single defect with the same total charge.

\section{Deconfined critical point in quantum antiferromagnets}
\label{SectionDQC}

In this section, we briefly review the critical theory for the N\'eel--VBS transition in spin-$\frac{1}{2}$ quantum antiferromagnets \cite{Senthil}. Consider the Hamiltonian
\beq{EqQuantumHamiltonian}
\scH = J \sum_{\langle r, r' \rangle} \Sv_{r} \cdot \Sv_{r'} + \scH_Q\punc{,}
\eeq
where \(J>0\), the sum is over neighboring pairs of sites $r$ and $r'$ of a 2D lattice, and $\Sv_{r}$ is a quantum-mechanical spin. The term $\scH_Q$ contains competing interactions, which can drive a zero-temperature transition into a VBS state, in which there is no magnetic order; an example is the \JQ\ model introduced by Sandvik \cite{SandvikJQ}.

When the first term in \refeq{EqQuantumHamiltonian} dominates, the ground state is a N\'eel antiferromagnet, with a two-sublattice collinear ordering and order parameter $\langle\Nv_{r}\rangle = \epsilon_r \langle \Sv_{r} \rangle \neq \zerov$, where $\epsilon_r = \pm 1$ on the two sublattices. When \(\scH_Q\) dominates, the ground state is instead a VBS, breaking the discrete lattice symmetries while preserving $\SU(2)$ symmetry. An order parameter for this phase is the (complex) expectation value of the VBS operator \(\psi\sub{VBS}\) \cite{Senthil}; different discrete ordering patterns are distinguished by its phase.

Studies of the square-lattice \JQ\ model using quantum Monte Carlo \cite{SandvikJQ,Block} give evidence for the claim of a direct continuous transition between N\'eel and VBS phases. This transition is believed \cite{Senthil} to be described by the same NC$CP^{1}$ critical theory  introduced in \refsec{SecCDM}. An argument for these claims is sketched below; readers are referred to \refcite{Senthil} for details.

A path-integral representation of a spin-\(\frac{1}{2}\) antiferromagnet can be expressed in terms of a unit vector $\hat{n}_r \sim \epsilon_r \Sv_r$. It contains, as well as terms corresponding to those in $\scH$, a Berry phase \cite{SachdevBook2} depending on the topology of the (periodic) path $\hat{n}_r(\tau)$ in imaginary time $\tau$; its effects will be addressed shortly. The path integral is more conveniently written in terms of spinons $\zm_r$, defined by \(\hat{n}_{r} = \zm_r^\dagger \sigmamv \zm_r\) with the constraint $\zm_r^\dagger \zm_r = 1$ required for normalization of the unit vector $\hat{n}_r$. This definition involves a gauge redundancy under phase rotation $\zm_r \rightarrow \zm_r \ee^{\ii\phi}$ at any site $r$, and so a long-wavelength theory also involves a \emph{compact} \(\U(1)\) gauge field \(\av\).

The N\'eel phase of the spin model, in which $\langle \hat{n} \rangle \neq \zerov$, is represented in these terms by an Anderson--Higgs phase, schematically expressed as a condensate of spinons, $\langle \zm_r \rangle \neq \zerom$. The Coulomb (deconfined) phase of the \NCP\ theory would correspond to a $\mathrm{U}(1)$ quantum spin liquid with deconfined spinons. In fact, no such spin-liquid phase exists in the spin model, because the compact nature of the gauge field \(\av\) allows for the existence of magnetic monopoles \cite{Polyakov}. In the absence of a spinon condensate (i.e., in a non-Higgs phase), monopoles are deconfined and replace the Coulomb phase by a ``monopole plasma'', with short-range correlations and no topological order. This phase, which is equivalent to the cubic dimer model in the disordered phase at nonzero fugacity, describes the VBS phase of the quantum antiferromagnet (despite the superficial similarity of the VBS to the dimer crystal).

To understand the latter identification, as well as the critical properties, it is necessary to consider the effect of the Berry phases in the path integral. As shown by Haldane \cite{Haldane}, the only important contributions to the global Berry phase are associated with ``hedgehog'' singularities of $\hat{n}$ in space-time or, equivalently, with monopoles. One can show \cite{Senthil} that an operator that inserts monopoles into the partition function has the same transformation properties under spatial symmetries as the VBS operator $\psi\sub{VBS}$. It follows that the long-distance limit of the ($q=1$) monopole distribution function vanishes in a spatially symmetric state, and conversely that a state with a nonzero limit must have broken spatial symmetry. These two statements imply, respectively, the following: (1) at a continuous transition into a VBS (at which point $\langle\psi\sub{VBS}\rangle = 0$), the monopole distribution function has a zero long-distance limit; and (2) in the ``monopole plasma'' on the non-N\'eel side of the transition, there must be VBS order.

One therefore has a scenario where single monopoles, though proliferating in the VBS phase, are absent at the transition. The same logic can be applied to monopoles of any charge $q$ for which the insertion operator transforms nontrivially under the lattice symmetries. Considering the symmetries of the VBS operator, one finds that this argument applies for all $q < q\sub{min}$, where $q\sub{min} = 4$ for the square lattice, $q\sub{min} = 3$ for honeycomb, and $q\sub{min} = 2$ for rectangular (twofold rotation symmetry). Monopoles of charge $q\sub{min}$ (and integer multiples thereof) are allowed at the critical point and will proliferate, leading to confinement of spinons, if the corresponding operators are relevant in the RG sense. If no such monopoles are relevant, the effective theory describing the transition will be noncompact, and given by the NC$CP^1$ critical theory introduced in \refsec{SecCDM}.

Available numerical evidence suggests that the \JQ\ model has a continuous transition on the square lattice but a first-order transition in the case of rectangular symmetry, implying that monopoles of charge $q=2$ are relevant at the NC$CP^1$ fixed point, while those with charge $q = 4$ are irrelevant \cite{Block}. For the honeycomb lattice, the picture is less clear \cite{Harada,Pujari,Pujari2}, and we therefore focus on the corresponding value of \(q = 3\).

\section{Scaling theory for monopoles}
\label{SecMethods}

\subsection{Monopole distribution function}
\label{SecMDF}

For a system of finite (linear) size \(L\), the scaling of the monopole distribution function \(G_q\), given in \refeq{EqGqLargeR}, can be extended to \cite{Cardy}
\beq{EqGqLargeRL}
G_q(\Rv,L) \approx L^{-2x_q} \Gamma_q(\Rv/L)\quad\text{for \(\lvert\Rv\rvert \gg a\)}\punc{,}
\eeq
where \(\Gamma_q\) is a universal function (for each \(q\)) and \(a\) is the lattice scale (set to unity elsewhere but made explicit in this section).

On the other hand, for \(\lvert\Rv\rvert\) much smaller than \(L\), whether larger than \(a\) or not, one expects
\beq{EqGqSmallR}
G_q(\Rv,L) \approx g_q(\Rv)\quad\text{for \(\lvert\Rv\rvert \ll L\)}\punc{,}
\eeq
independent of \(L\). This follows from the definition of \(G_q\) in terms of a charge-neutral, and hence topologically trivial, ensemble, whose free energy relative to the zero-monopole ensemble is independent of \(L\).

Combining \refeqand{EqGqLargeRL}{EqGqSmallR} in the intermediate regime \(a \ll \lvert\Rv\rvert \ll L\) gives
\beq{EqGammaRho}
\Gamma_q(\rhov) \sim \lvert\rhov\rvert^{-2x_q}\quad\text{for \(\lvert\rhov\rvert \ll 1\).}
\eeq
Together with \refeq{EqGqLargeRL}, this simply states that the monopole distribution function is a power law in \(\lvert\Rv\rvert\) in the thermodynamic limit.

The Markov-chain MC method can give the ratio of partition functions for two charge configurations \(\{Q_{\rv}\}\), provided that one can transition between the two with rates respecting detailed balance. We are therefore able to determine the ratio
\beq{EqDefinescG}
\scG_q(\Rv,\Rv';L) = \frac{G_q(\Rv,L)}{G_q(\Rv',L)}
\eeq
by implementing a MC update that allows charge-\(q\) monopoles to move, described in \refsecand{SecCharge2updates}{SecCharge3updates}.

A straightforward way to extract \(x_q\) from MC results for \(\scG_q\) is to choose fixed \(\Rv/L\) and \(\Rv'/a\), both of order unity \cite{MonopoleScalingPRB,Sreejith}. Their ratio then scales as
\beq{EqscGscaling1}
\scG_q(\Rv,\Rv';L) \approx \frac{L^{-2x_q}\Gamma_q(\Rv/L)}{g_q(\Rv')}\punc{,}
\eeq
allowing \(x_q\) to be determined through finite-size scaling. Plots of \(L^{2x_q}\scG_q(\Rv,\Rv';L)\) versus \(\rhov = \Rv/L\) for different \(L\) (with fixed \(\Rv'\)) should collapse onto a single curve (for each \(q\)), proportional to the universal function \(\Gamma_q\).

An alternative method is to find \(\scG(b\Rv, \Rv;L)\) for \(b\) greater than but of order unity. For intermediate separations, such that \(a \ll \lvert\Rv\rvert \ll L/b\), one finds, using \refeqand{EqGqLargeRL}{EqGammaRho},
\beq{EqScalingFixedL}
\scG(b\Rv, \Rv;L) \approx \frac{\Gamma_q(b\Rv/L)}{\Gamma_q(\Rv/L)} \approx b^{-2x_q}\punc{.}
\eeq
This form may be more reliable in cases where finite-size scaling is problematic, as reported in  \refcite{Nahum2}.

\subsection{Additional interactions between monopoles}
\label{AdditionalInt}

In the cases of most interest, where \(y_q = d - x_q\) is small, the ratio in \refeq{EqscGscaling1} decreases rapidly with \(L\) and the relative statistical error increases. To improve convergence, one can add an explicit interaction between monopoles, by defining
\beq{EqDefineZQ2}
\scZ_L[Q_{\rv},\Phi(\Rv)] = \ee^{-\sum_{\rv,\rv'}'Q_{\rv}Q_{\rv'}\ln\Phi_L(\rv-\rv')} \sum_{\hsmash{\substack{\{B_\mu(\rv)\}\\ \Div_{\rv} B = Q_{\rv}}}} \ee^{-E\sub{dimer}/T}\punc{,}
\eeq
where the summation \(\sum_{\rv,\rv'}'\) excludes \(\rv=\rv'\) and each pair of sites is counted only once. The function \(\Phi_L\) should be symmetric and should depend on \(L\) such that it respects the periodic boundaries, but is otherwise arbitrary. Ratios of partition functions can be calculated by including the interaction \(\Phi_L\) within the MC weights. (Note that \(\Phi_L > 1\) increases the Boltzmann weight of a configuration with oppositely charged monopoles.)

The quantity \(\scG_q(\Rv,\Rv';L;\Phi_L)\) can be defined by analogy to \refeq{EqDefinescG}, and is given by
\beq{EqscGPhi}
\scG_q(\Rv,\Rv';L;\Phi_L) = \left(\frac{\Phi_L(\Rv)}{\Phi_L(\Rv')}\right)^{q^2}\scG_q(\Rv,\Rv';L)\punc{.}
\eeq
If one fixes \(\rhov = \Rv/L\) and \(\Rv'/a\) of order unity and chooses a function \(\Phi_L\) such that \(\Phi_L(\rhov L)\approx cL^{2\theta/q^2}\) (for fixed \(\rhov\)) and \(\Phi_L(\Rv')\approx c'\lvert\Rv'\rvert^{2\theta/q^2}\) (independent of \(L\)) for some $\theta$, then one finds
\beq{EqscGPhiScaling}
\scG_q(\Rv,\Rv';L;\Phi_L) \approx \frac{L^{2(\theta-d+y_q)}\Gamma_q(\Rv/L)}{g_q(\Rv')\lvert\Rv'\rvert^{2d}}\left(\frac{c}{c'}\right)^{q^2}\punc{.}
\eeq
Scaling with \(L\) then gives access to the RG eigenvalue \(y_q\). The intermediate functional form of \(\Phi_L\) should be chosen so that the distribution is reasonably flat, to maintain ergodicity and efficiency of the algorithm.

A possible choice of additional interaction is given, for sufficiently large \(\theta\), by
\beq{EqPhiL}
\Phi_L(\Rv) = \left( \sum_{\Tv \in \operatorname{pbc}(L)} \left\lvert\Rv-\Tv\right\rvert^{-2\theta} \right)^{-1/q^2}\punc{,}
\eeq
where \(\operatorname{pbc}(L)\) is the set of all vectors \(\Tv\) such that translation by \(\Tv\) is equivalent to the identity transformation under the boundary conditions. For \(\lvert\Rv\rvert \ll L\), the term with \(\Tv = \zerov\) dominates the sum and \(\Phi_L(\Rv)\approx \lvert\Rv\rvert^{2\theta/q^2}\), while for \(\Rv = \rhov L\) with \(\lvert\rhov\rvert\) fixed and of order unity, \(\Phi_L(\Rv) \propto L^{2\theta/q^2}\).

\section{Numerical Methods}
\label{SecNumericalMethods}

\begin{figure}
\includegraphics[width=\columnwidth]{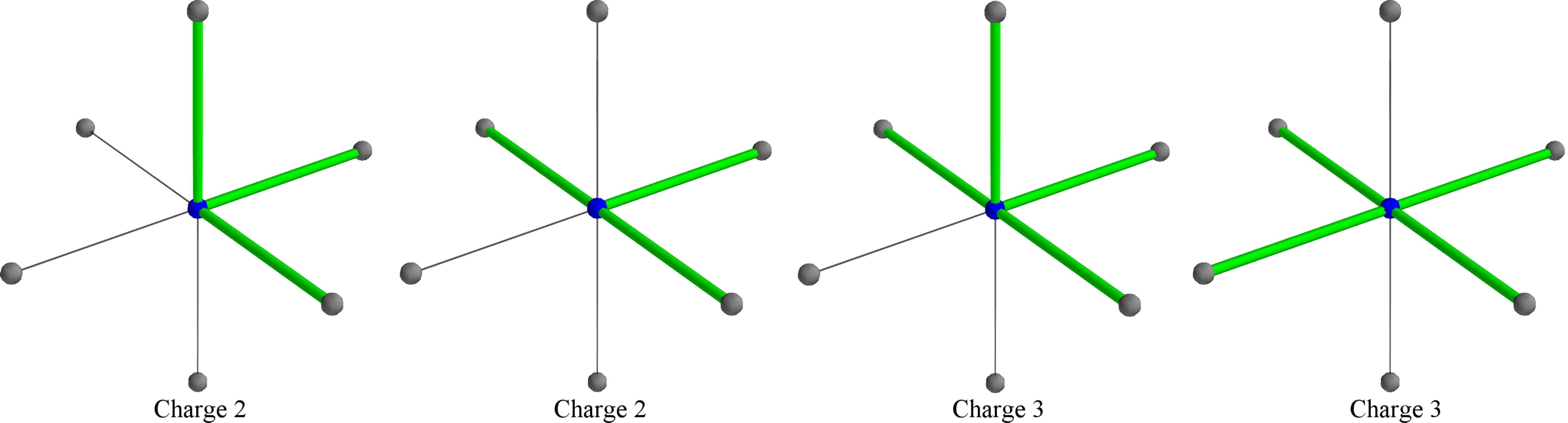}
\caption{Illustration of a local charge $q$, consisting of a site with $q+1$ intersecting dimers. Charge \(q = 2\) and \(3\) occur in the configurations shown above and their rotations.}
\label{ChargesIllustration23}
\end{figure}

This section describes the numerical methods used to calculate the distribution function \(G_q\) for a pair of monopoles of charge $\pm q$ at the critical point (for \(q=2\) or \(3\)).
The allowed configurations in the constrained cubic dimer model are those where every site has exactly one dimer connected to it. In order to study the correlation function of a pair of charges, we consider a new set of  configurations, but now with the dimer constraints violated at any two sites $\rv_1$ and $\rv_2$, located on opposite sublattices $A$ and $B$. As illustrated in \reffig{ChargesIllustration23}, $q+1$ dimers overlap at these sites, creating a charge of $\pm q$, according to \refeq{EqDivB}. The configuration space
\begin{equation}
\scC_q=\bigcup_{\hsmash{\substack{\rv_1\in A\\\rv_2\in B}}} \scC_q(\rv_1,\rv_2)
\end{equation}
thus contains all possible dimer configurations, with all possible locations \(\rv_{1,2}\) of the charges. We use MC methods to sample with Boltzmann weight $\propto \exp\left({-{E}/{T}}\right)$ over $\scC_q$, where the energy $E$ of a configuration is given by \refeq{EqDefineE}. Given such samples, the pair distribution function is calculated using
\begin{equation}
G_q(\rv_1,\rv_2) \propto  \frac{\text{number of samples in \(\scC_q(\rv_1,\rv_2)\)}}{\text{number of samples in \(\scC_q\)}}\punc{.}
\end{equation}

In order to sample from the full configuration space, we employ two different MC update steps, both of which satisfy detailed balance. The first update process, $\scT_1$, changes the configuration of the dimers but preserves the locations of the charges. The second update process, $\scT_2$, moves one of the two charges to a nearby point on the same sublattice. Details of the updates are given in \refsecs{SecBackgroundUpdates}{SecCharge3updates}. At each MC step, denoted \(\scT\), either $\scT_1$ or $\scT_2$ is chosen with equal probability. Thermalization from an initial ordered state is assumed to have been achieved after attempting $1000\times L^3$ dimer flips using $\scT_1$ and $\scT_2$. After thermalization, the average number of update steps $\mathcal{N}_{L^3}$ required for $L^3$ dimer flips is estimated by averaging over several update steps. Having estimated $\mathcal{N}_{L^3}$, samples are then taken once every $\mathcal{N}_{L^3}$ update steps $\scT$. Multiple runs give independent estimates of $G_q$ allowing error estimation using the jack-knife method.

\subsection{Updates of the background dimer configurations}
\label{SecBackgroundUpdates}

Here we describe the MC steps $\scT_1$ which sample within the subset $\scC_q(\rv_1,\rv_2)$ with the charges located at fixed sites $\rv_{1,2}$. The method is based on the directed-loop algorithm \cite{SandvikMoessner}, and the specific implementation is very similar to the one used in \refcite{Sreejith}. Here we give a summary for completeness.

\begin{figure}
\includegraphics[width=\columnwidth]{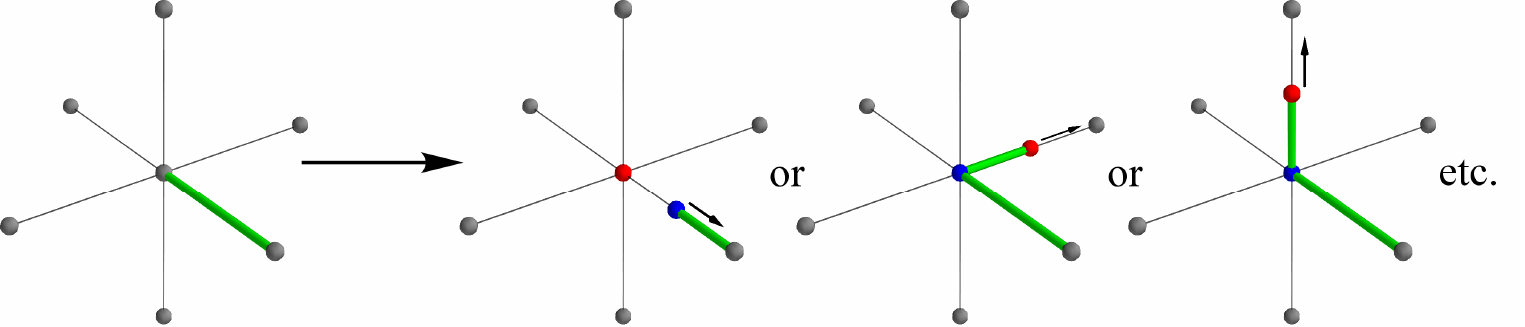}
\caption{Initial step of the update process for background dimers. The link-monomer and the monomer at the starting node $\rv_0$ are indicated by red/blue dots. The arrows in the figure indicate the direction of the link monomer.}\label{initialstepfig}
\end{figure}

The update starts by choosing a starting node $\rv_0$, from any of the charge-free lattice sites, with equal probability. The node is accepted with a probability that depends on the local configuration (described below). If the step is accepted, the system transitions into an intermediate configuration in which there is a charge $\pm 1$ (monomer) located at \(\rv_0\) and another one of charge $\mp 1$ on a link connected to it. Creation of two such charges is accompanied by addition or removal of half a dimer as shown in the example in \reffig{initialstepfig}. The link monomer has a `direction' associated with it, which initially, is set to be away from $\rv_0$. The direction identifies one of the two nodes in that link as the node `ahead' of the link monomer.

\begin{figure}
\includegraphics[width=\columnwidth]{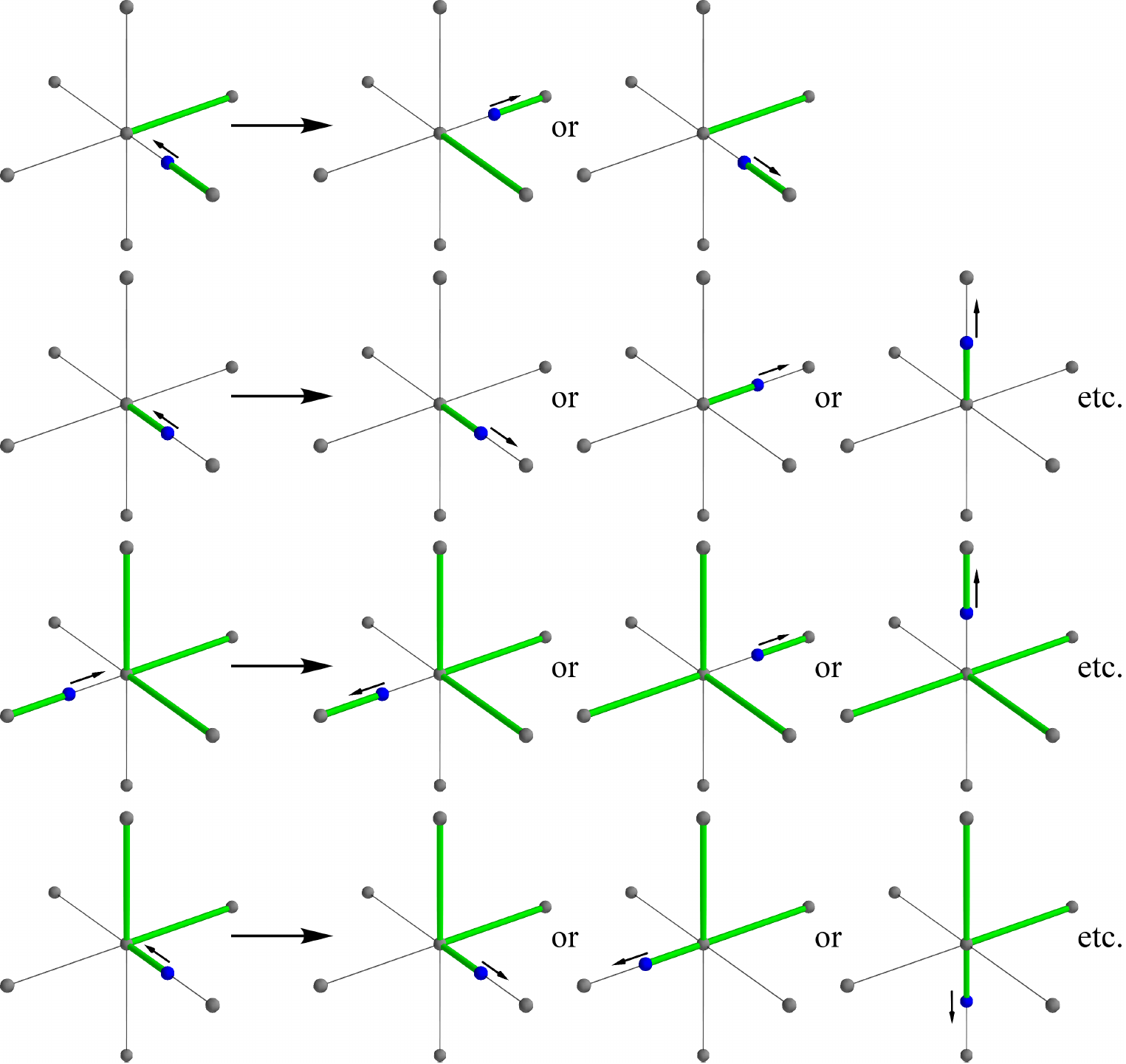}
\caption{Examples of individual hopping steps of the link monomer (blue dot). The first two rows show the scenario when the node ahead has no charge. The third and fourth lines show the step where the node ahead has a charge $2$. Only hopping steps that do not change the local charge are allowed. These steps can be generalized to the case where the node has higher $q$.}\label{propogationillustration}
\end{figure}

\begin{figure}
\includegraphics[width=\columnwidth]{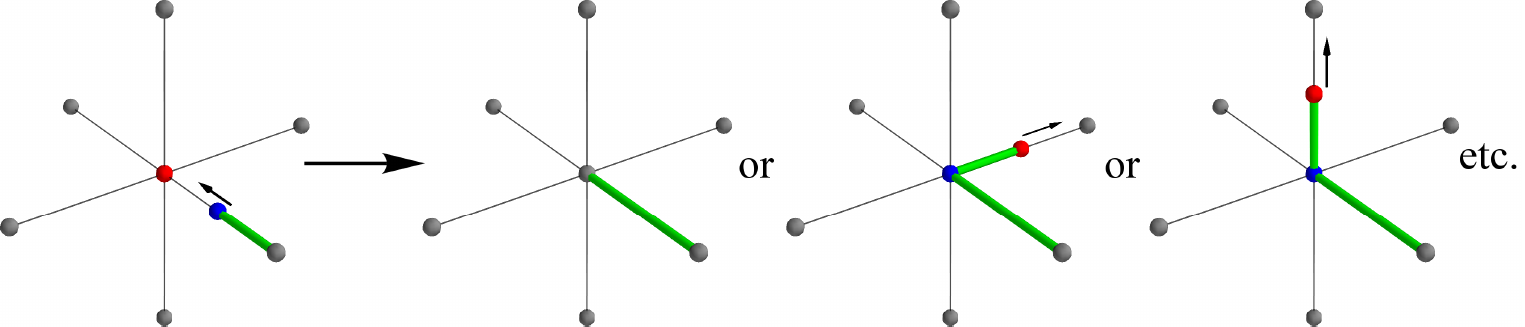}
\caption{An example of a termination step for the background dimer update. When the link monomer returns to the starting point \(\rv_0\) of the update loop, it can annihilate the monomer on the starting site and result in a new configuration in $\scC_q(\rv_1,\rv_2)$.}\label{terminationfig}
\end{figure}

In further steps, the link monomer can hop to one of the six links connected to the node $\rv_i$ ahead of it, erasing or creating dimers in the process, as shown in \reffig{propogationillustration}.
The direction of the link monomer after the hop is set to be away from $\rv_i$. Any move that results in a change in the charge at $\rv_i$ is not considered, unless $\rv_i=\rv_0$. In this case, the link monomer can also annihilate the monomer at $\rv_0$, as illustrated in \reffig{terminationfig}, completing the update step and giving a new configuration in $\scC_q(\rv_1,\rv_2)$.

When the link monomer hops over a site with a charge $q$ (see \reffig{propogationillustration}), the $q+1$ dimers composing the charge are rearranged, but the location and charge $q$ are unchanged.

The probability of transitions at each individual hop of the link monomer, as well as of the starting and terminating steps, are given by 
\begin{equation}
p(k\to q) = \frac{\bar{w}_q - \delta_{kq} \min(\bar{w})}{\sum \bar{w} - \min(\bar{w})}\punc{,}
\end{equation}
where $k$ and $q$ represents the initial and final configurations. The weight of a configuration \(q\) is given by $\bar{w}_q=\exp(-{\bar{E}_q}/{T})$, where $\bar{E}_q$ is calculated by taking the interaction energy of a half dimer to be half that of a full dimer. The notations $\min(\bar{w})$ and $\sum \bar{w}$ represent the minimum and the sum of the weights associated with all allowed final configurations $q$. This choice of probabilities ensures that detailed balance is satisfied for a complete step \(\scT_1\).

Note that in order to evaluate $p(k\to q)$ only the energy differences between configurations need to be calculated. These energy differences can be calculated by accessing the dimer occupancy in a very small portion of the whole system.

\subsection{Updates for charge-2 monopole}
\label{SecCharge2updates}

\begin{figure*}
\includegraphics[width=0.8\columnwidth]{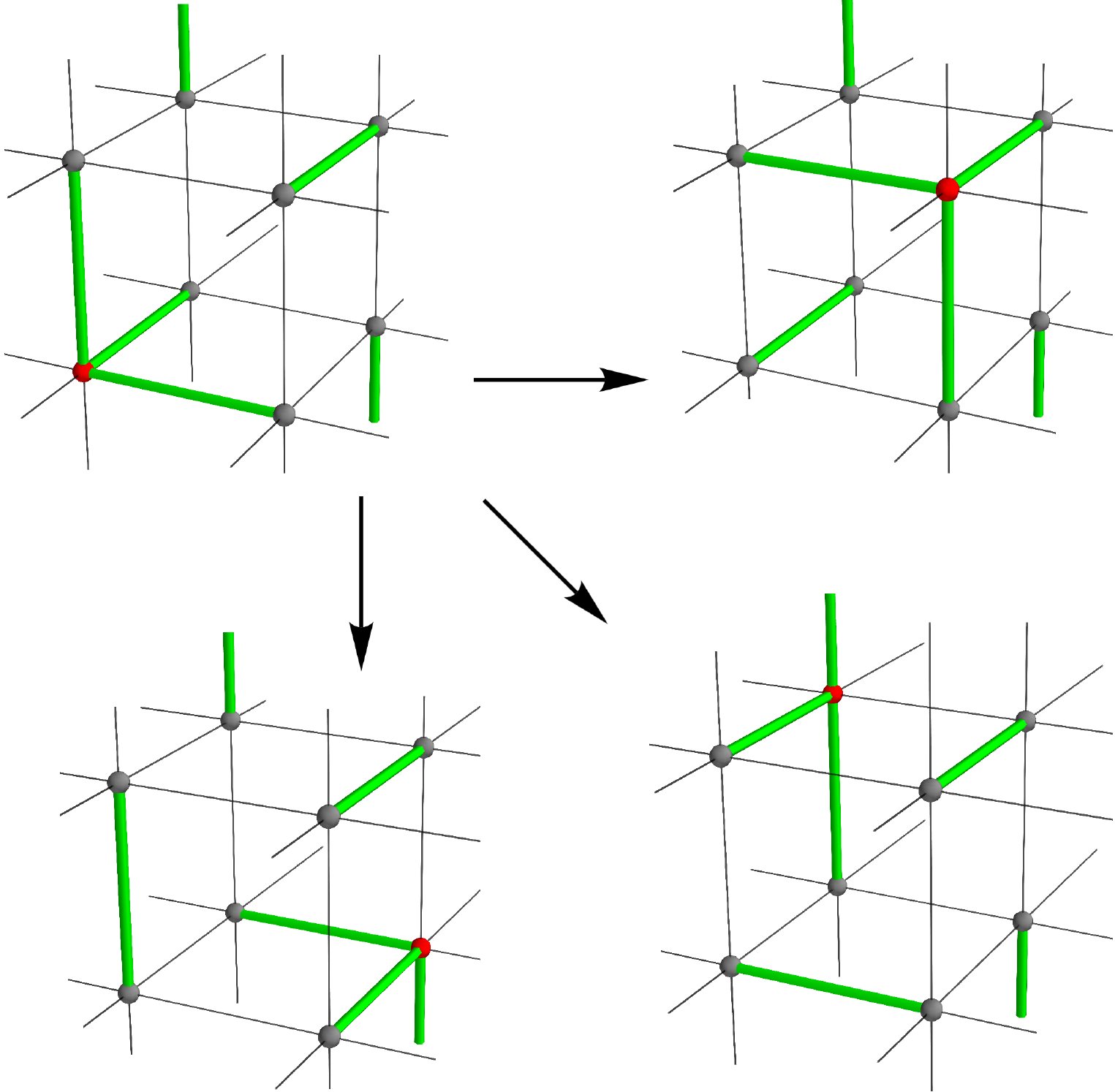}\hspace{4em}
\includegraphics[width=\columnwidth]{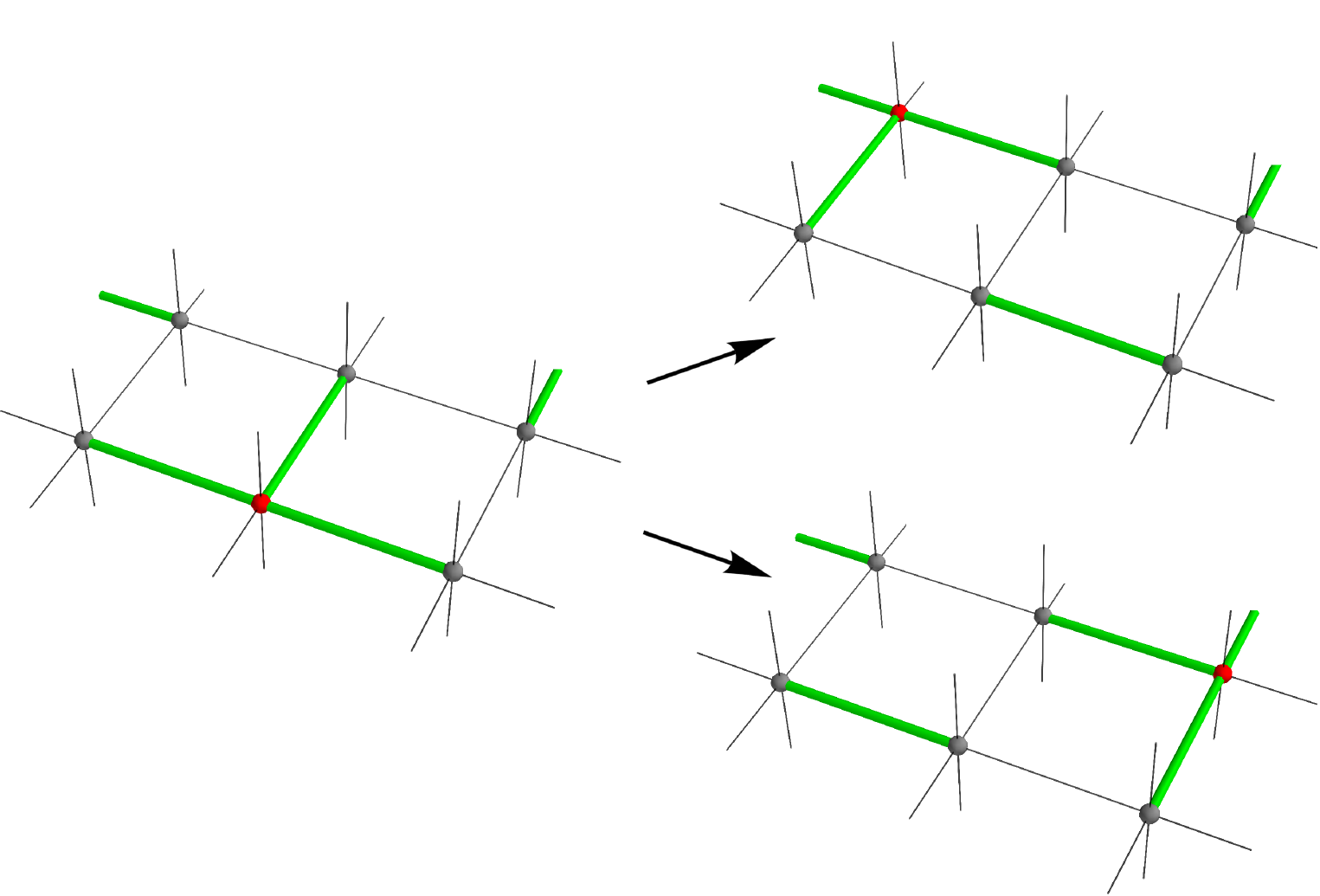}
\caption{Examples of the update step that allow a charge-$2$ monopole to be moved. The location of the charge is marked with red dot. Each possible translation of the charge involves moving exactly two of the three dimers constituting the monopole. Depending on the orientation of the dimers at the charge, the number of ways to hop out of the initial state can differ.}
\label{charge2hopping}
\end{figure*}

Here we describe the update steps $\scT_2$ used for achieving transitions between configurations with different locations of the $q=2$ charges. At each step of the update one of the two charges can move to one of the nearest points on the same sublattice through local rearrangement of dimers, as shown in \reffig{charge2hopping}.

In order to sample the distribution \(\scC_q\), the probabilities of the transitions need to be chosen correctly. This can be achieved using an acceptance--rejection method, but the different number of possible moves associated with different charge configurations means that the standard Metropolis probabilities must be modified, as follows: At the beginning of each step, one of the two charges is chosen at random (with equal probabilities). Let $n_0$ be the number of possible ways in which the selected charge can hop. One of these $n_0$ transitions is chosen at random (with equal probabilities). The selected transition is accepted with probability $\min{(1,\frac{w_1 n_0}{w_0 n_1})}$, where $n_1$ is the number of possible ways the charge can hop out of its new location, and $w_{0,1}$ are the Boltzmann weights of the old and new configurations.

\begin{figure}
\includegraphics[width=0.9\columnwidth]{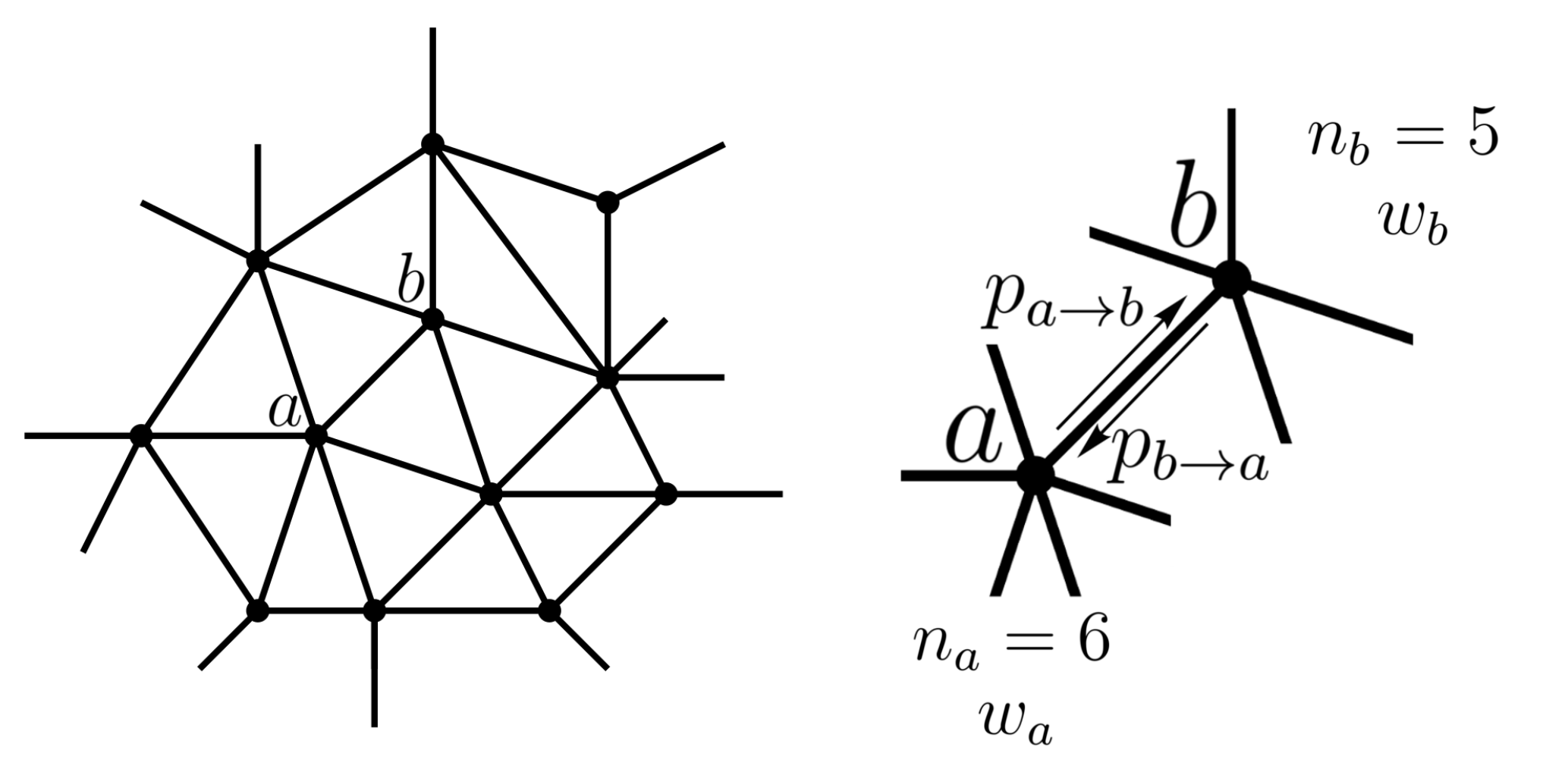}
\caption{The MC procedure can be thought of as forming a graph with edges connecting the configurations that are related by a single $\scT_2$ MC step. Thus transition probabilities $p_{a\to b}$ and $p_{b\to a}$ are nonzero iff they are adjacent on this graph. The number of transitions out of (or into) a configuration $a$ gives the degree $n_a$. Transition probabilities need to be picked such that configurations are sampled with the desired Boltzmann probabilities.}\label{graph}
\end{figure}

To clarify the choice of  probabilities, consider an initial configuration $a$ with Boltzmann weight $w_a$. Given a choice of one the two charges, one of the $n_a$ possible transitions is first chosen with probability $\frac{1}{n_a}$; see \reffig{graph}. This transition is accepted with probability $\frac{w_b n_a}{w_a n_b}$ (supposing ${w_b n_a} < {w_a n_b}$). The net effective probability of transition is thus $p_{a\to b}=\frac{1}{n_a}\times \frac{w_b n_a}{w_a n_b}$, while, for the reverse process, the effective probability is $p_{b\to a}=\frac{1}{n_b}\times 1$. These satisfy detailed balance $w_a p_{a\to b}=w_b p_{b\to a}$.

Note that the number of ways $n_0$ a given charge can hop depends on the configuration of the $q+1$ dimers connected to the charge. When the three dimers forming the charge (for \(q = 2\)) are coplanar, there are two ways in which the charge can hop to a new point. If instead the dimers are non-coplanar, the charge can hop in three different directions (see \reffig{charge2hopping}). In addition, when the two charges are close to each other, the dimers attached to one charge can block some of the transitions out of the current state, thus affecting the number of transitions for a given charge.

\subsection{Updates for charge-3 monopole}
\label{SecCharge3updates}

\begin{figure}
\includegraphics[width=0.8\columnwidth]{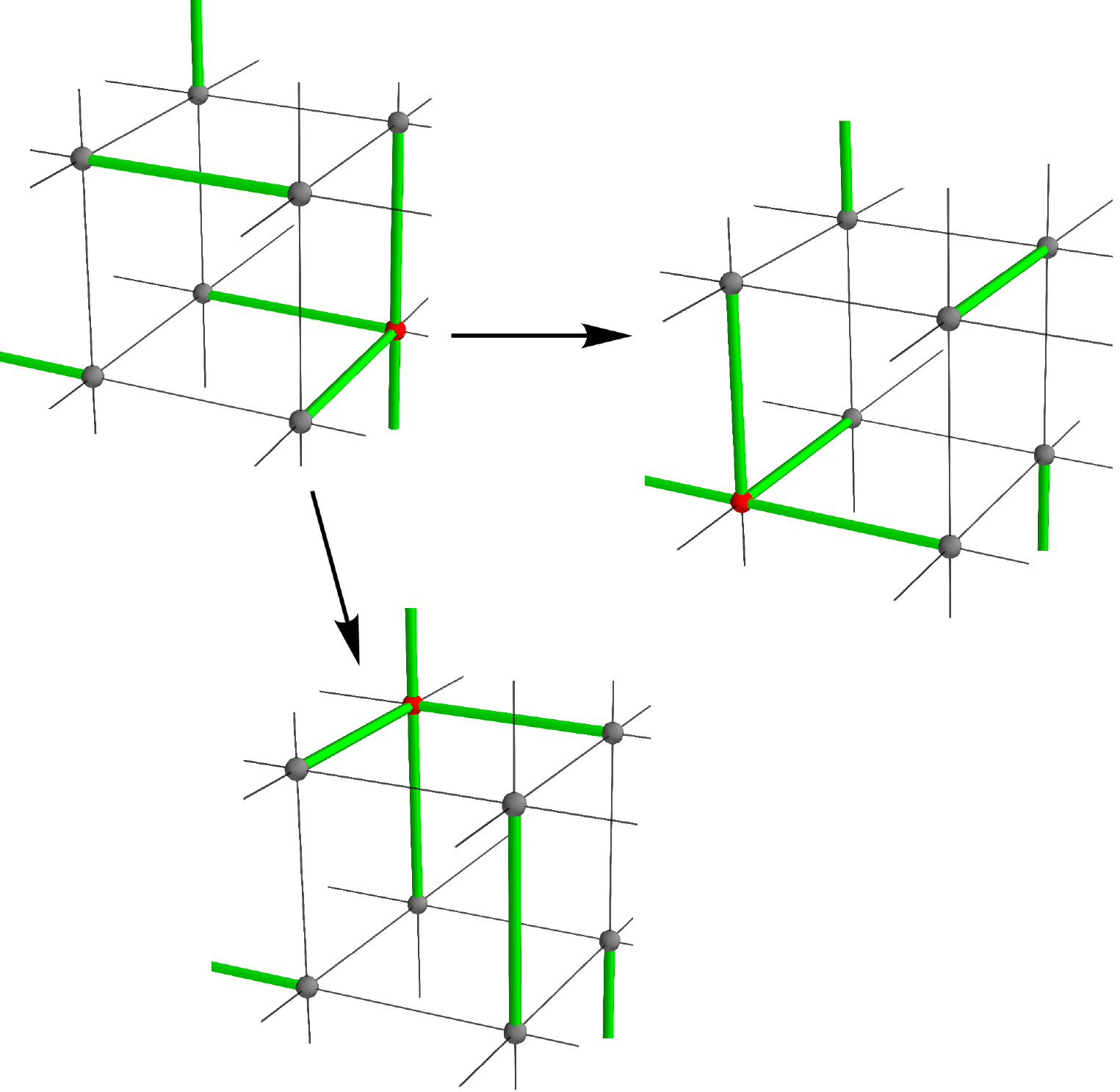}
\caption{Examples of the update step that allow a charge-3 monopole to move. When the dimers at the charge $q=3$ are in a non-coplanar configuration, three out of the four dimers lie on adjacent edges of a unit cube. There are two such triplets and two associated cubes. When there is a dimer at an edge diagonally opposite to any of these three dimers, as shown above, there are two possible transitions of the charge out of the configuration. Note that such transitions always result in a non-coplanar arrangement of dimers.}
\label{charge3transition}
\end{figure}

The update steps \(\scT_2\) for transitions between configurations with different locations of the charge-3 monopole are similar to those described for charge \(2\). There are zero possible transitions into or out of the coplanar $q=3$ dimer configuration shown in fourth panel of \reffig{ChargesIllustration23}. 
When the dimers are not coplanar, (third panel of \reffig{ChargesIllustration23}), there are two or four transitions possible as described here. In this configuration, three out of the four dimers lie on mutually perpendicular directions. There are two such triplets among the four dimers. Dimers in each such triplet lie along three adjacent edges defining a cube as shown in  \reffig{charge3transition}. If there is a dimer occupying an edge diagonally opposite to any of the three dimers, the dimers within the cube can be rearranged in two different ways as shown in the example. Each such rearrangement moves the charge to a new location. Since there are two different triplets and associated cubes, there can be two or four transitions in total. Similar to the case of charge $2$, the number of transitions out of a charge are again modified when the two charges are in close proximity.

Note that, since there are no transitions of the type $\scT_2$ that can move a charge starting from a coplanar configuration, one would expect (from detailed balance) that there are no $\scT_2$ transitions into a coplanar configuration; this is in fact true for the scheme described. Ergodicity is maintained, however, because a  $\scT_1$ update step can connect a coplanar configuration and a non-coplanar configuration.

\subsection{Improving convergence}
\label{SecImprovingConvergence}

As noted in \refsec{AdditionalInt}, the correlation function decays rapidly with \(\lvert\Rv\rvert\) and a large number of iterations is therefore needed to obtain accurate estimates. A sample at a large distance $R\sim \mathcal{O}(L)$ is obtained only once in $\mathcal{O}(L^{2x_{q}})$ steps. We have used two methods for improving convergence for the charge $3$ systems.

\subsubsection{Piecewise estimation}
\label{SecPiecewise}

One way to reduce the computational time is through estimating the correlation functions in two segments. The two correlation functions $\mathcal{G}_{<}(R,R'=1,L)$ and $\mathcal{G}_{>}(R,R'=\frac{2}{3}R_{\rm max},L)$ are calculated in separate MC runs by restricting the samples to
\begin{align}
\scC_< &=\scC_q\left(0<|\rv_1-\rv_2|<\tfrac{2}{3}R_{\rm max}\right)\\
\scC_> &=\scC_q\left(\tfrac{1}{3}R_{\rm max}<|\rv_1-\rv_2|<R_{\rm max}\right)\punc{,}
\end{align}
respectively, where $R_{\rm max}\simeq\frac{\sqrt{3}}{2}L$ is the maximum possible distance between the charges. This is in practice achieved by rejecting any transition $\scT_2$ that takes the system out of the subsets.

The two results can be patched together at $R=\frac{1}{2}R_{\rm max}$ to obtain the full function $\scG_q$. Note that there is no constraint on the relative locations of charge-one monomers that occur in the directed-loop algorithm, step \(\scT_1\).  We have checked in systems of sizes up to $L=30$ that the correlation functions obtained this way are identical to that obtained with full MC sampling in the entire region.

\subsubsection{Repulsive interaction between charges}
\label{SecConvergence}

Another way to improve the estimation is to use an added repulsive potential, as explained in \refsec{AdditionalInt}. Samples are obtained from the Boltzmann distribution where the energy functional has an added repulsive interaction energy between the charges of the form of \refeq{EqPhiL}. This interaction term has a sum over the repulsion from mirror charges in copies of the whole lattice system arising from periodic boundary conditions. Computationally this interaction is calculated by summing over $20$ periodic lattices, corresponding to $-20L\leq T_i\leq 20L$ for $i=1,2,3$ in \refeq{EqPhiL}. The specific results presented here for charge $3$ were estimated with $\theta=\frac{5}{2}$. Estimates made with the added repulsive interaction match the values obtained without it within error bars.

\section{Results}
\label{SecResults}

In the following subsections we describe our results for the monopole distribution function \(G_q\) for charges $q=2$ and $3$.

\subsection{Charge 2}
\label{SecResultsCharge2}

\begin{figure}
\includegraphics[width=\columnwidth]{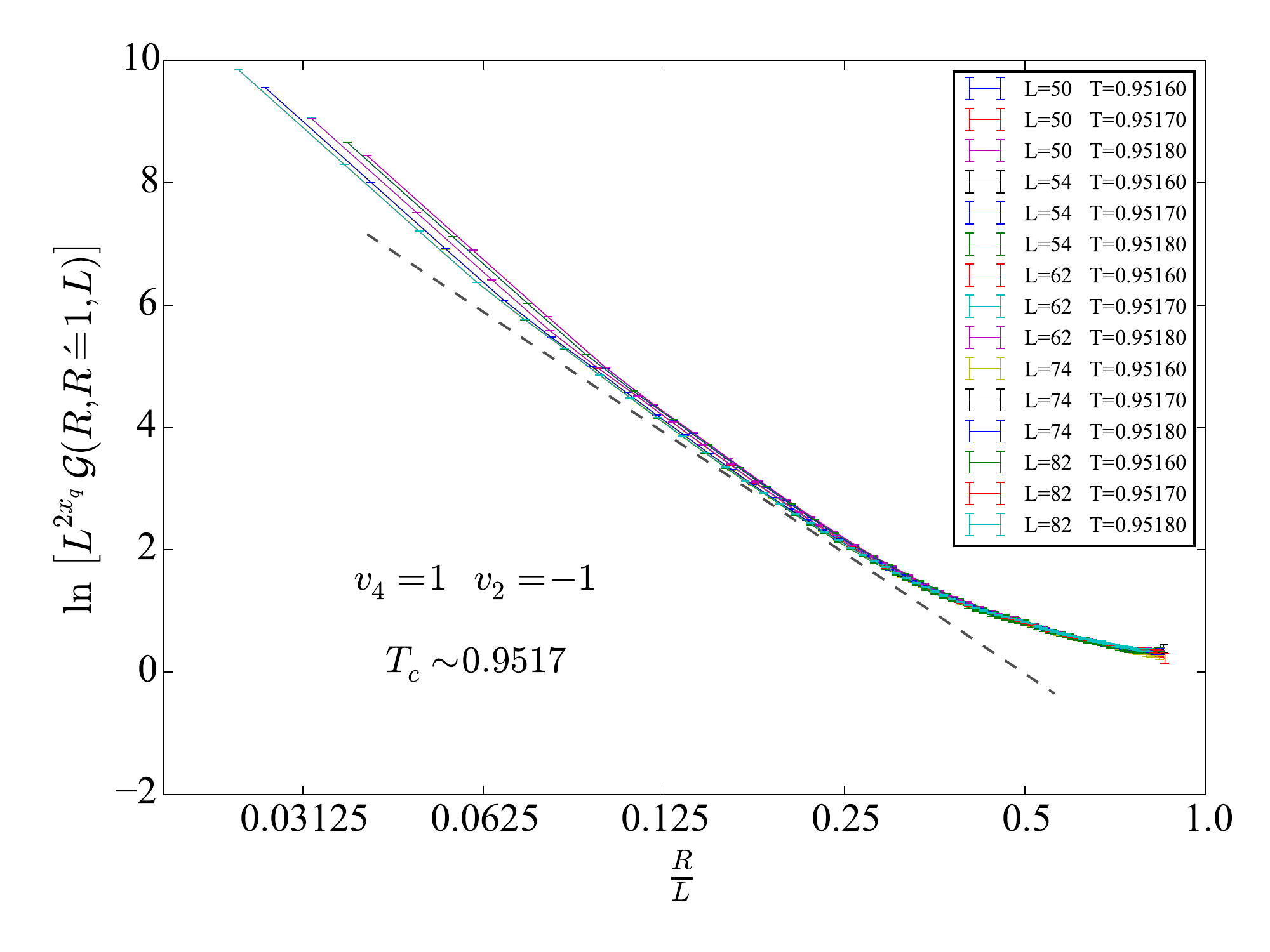}
\includegraphics[width=\columnwidth]{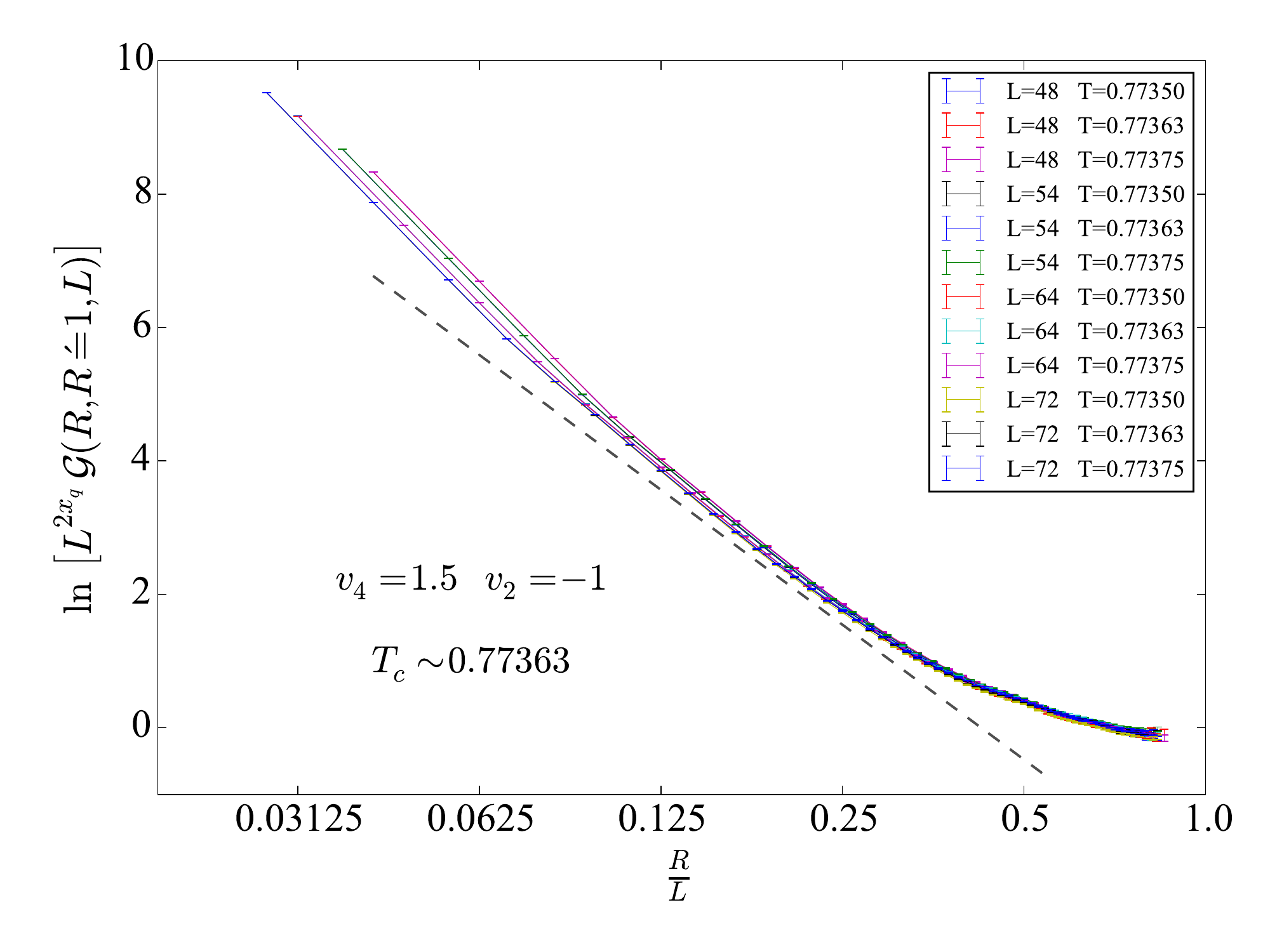}
\includegraphics[width=\columnwidth]{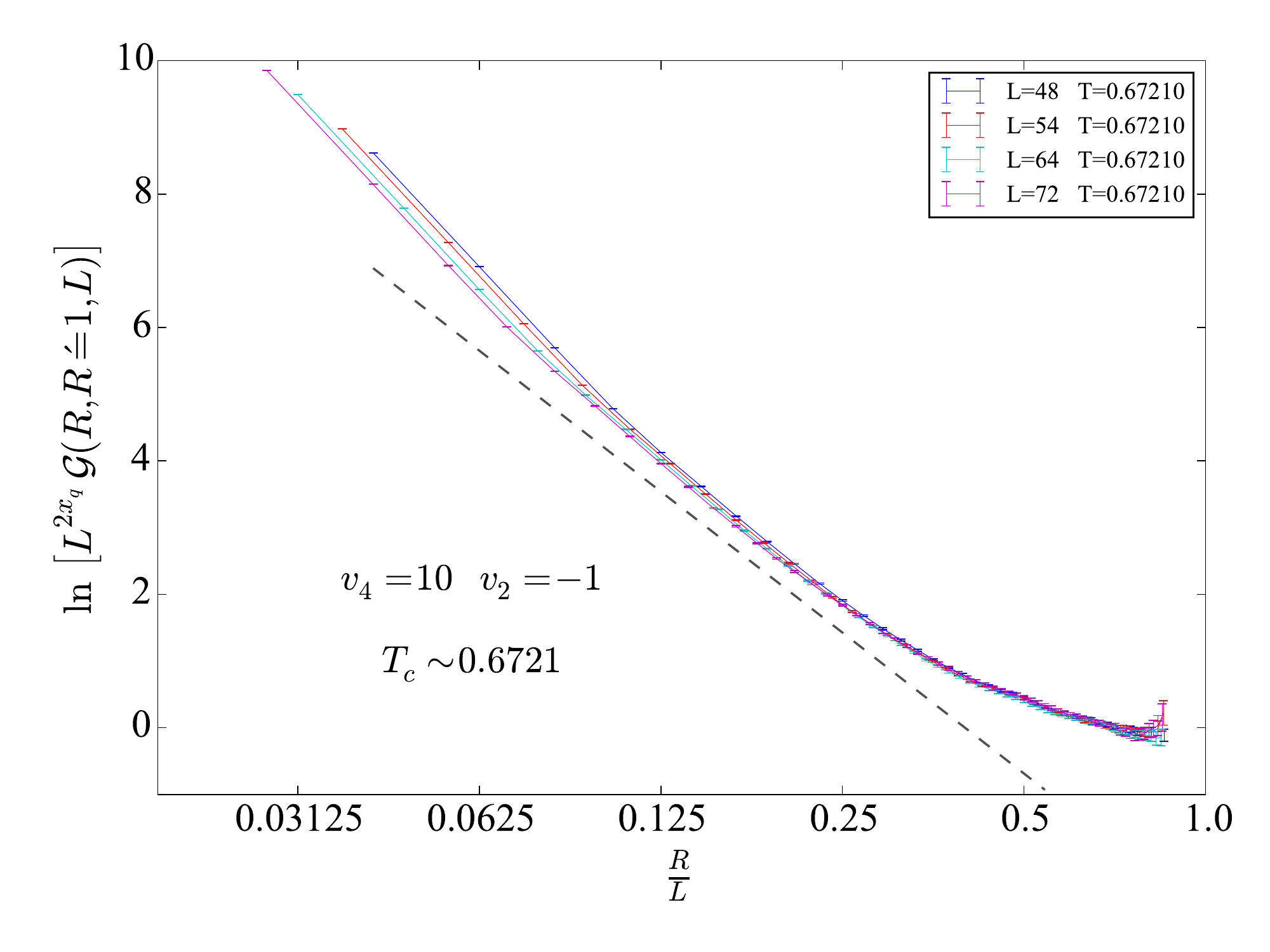}
\caption{Monopole distribution function \(\scG_q\) for a pair of charges with \(q = \pm 2\). The two figures show the data for different four-dimer interactions $v_4=1.0$ (top), $1.5$ (middle), and $10$ (bottom). The functions have been rescaled with $L^{2 x_q}$ where the scaling dimension \(x_q\) is related to the RG eigenvalue \(y_q\) obtained from \reffig{scaling_charge_2} by $x_q=d-y_q$. The dashed lines have slope $-2x_q$, and are expected to be parallel to the scaled correlation for $1\ll R\ll L$.}
\label{Q2pd}
\end{figure}
\begin{figure}
\includegraphics[width=\columnwidth]{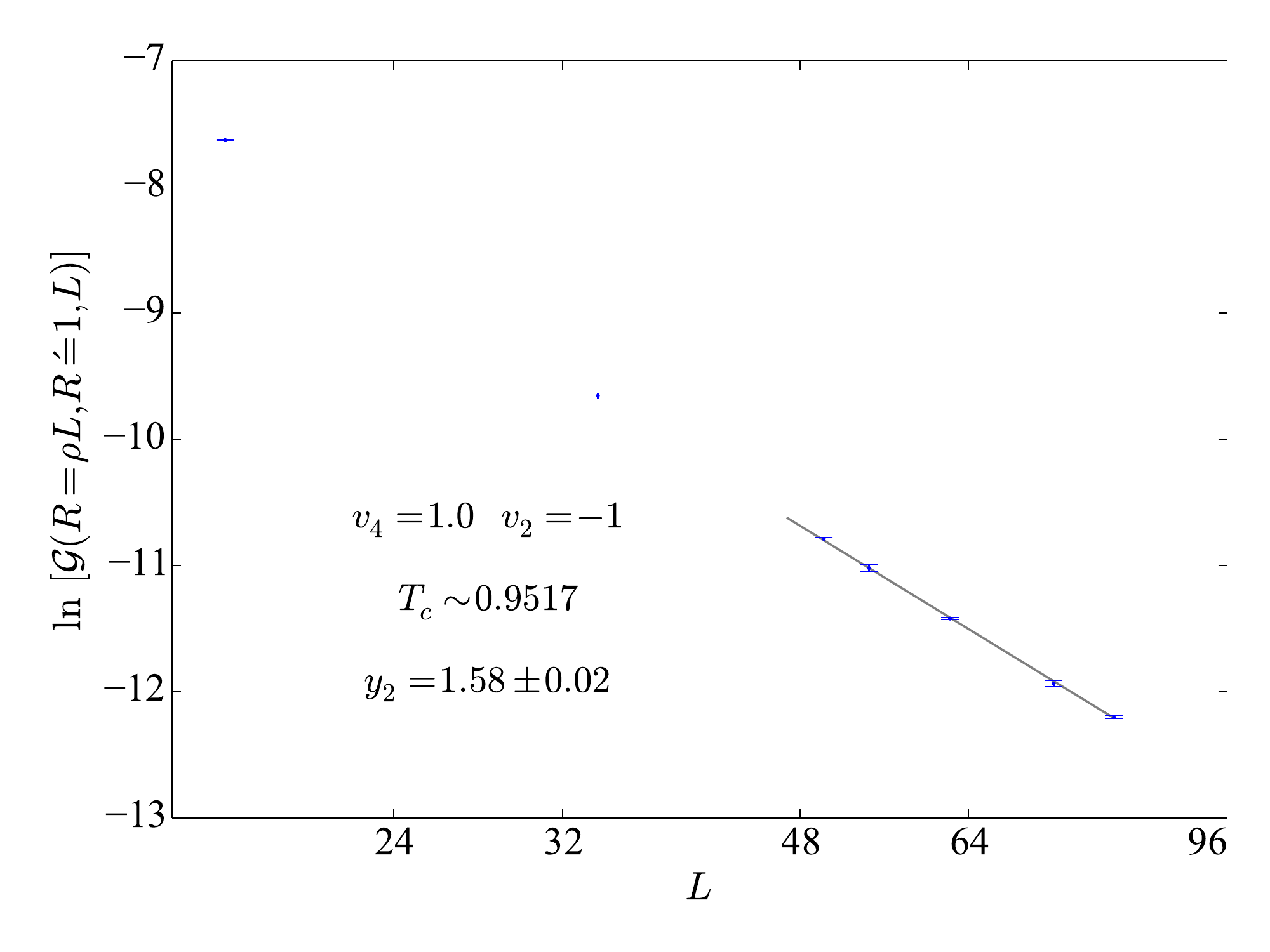}
\includegraphics[width=\columnwidth]{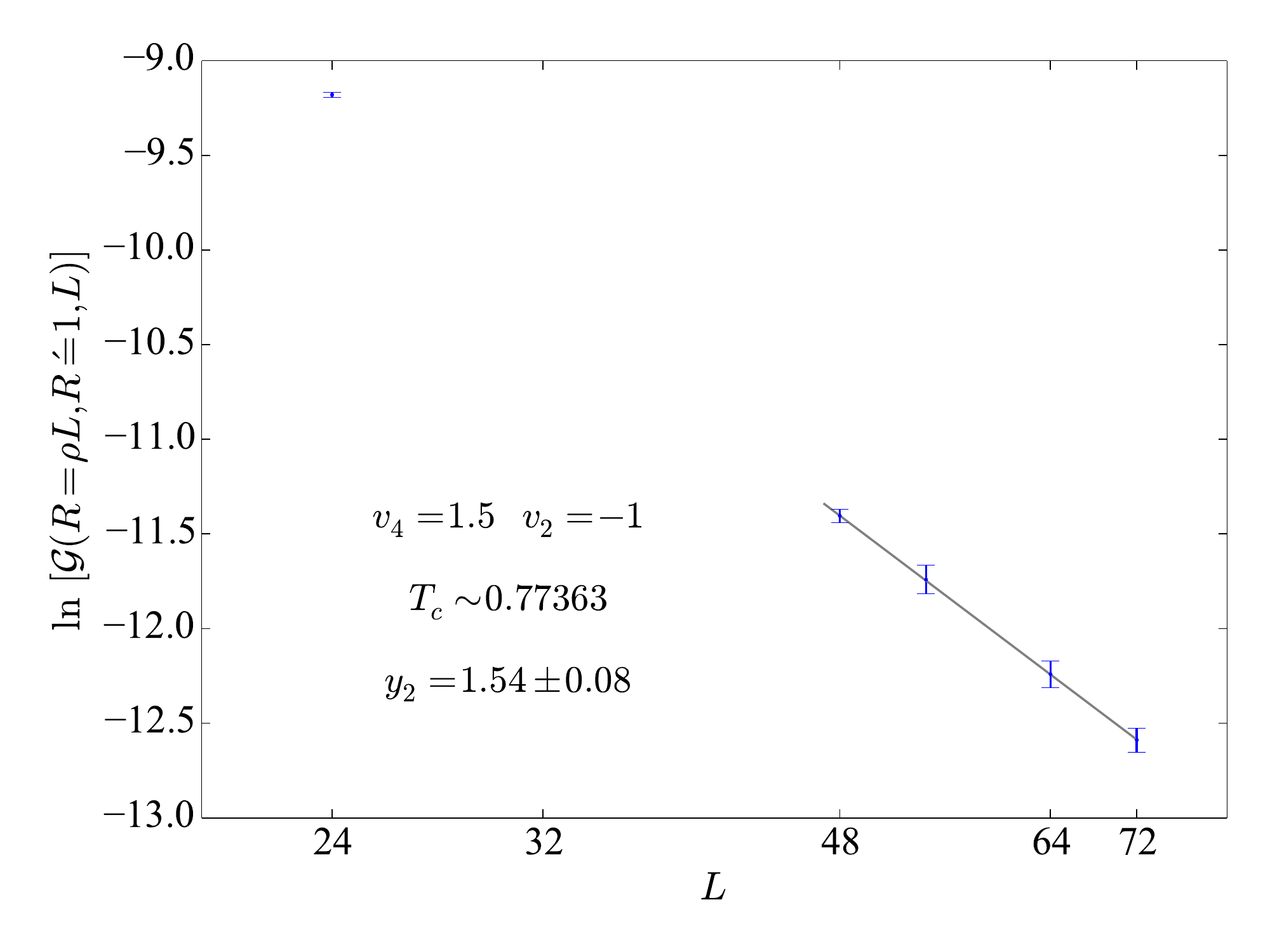}
\includegraphics[width=\columnwidth]{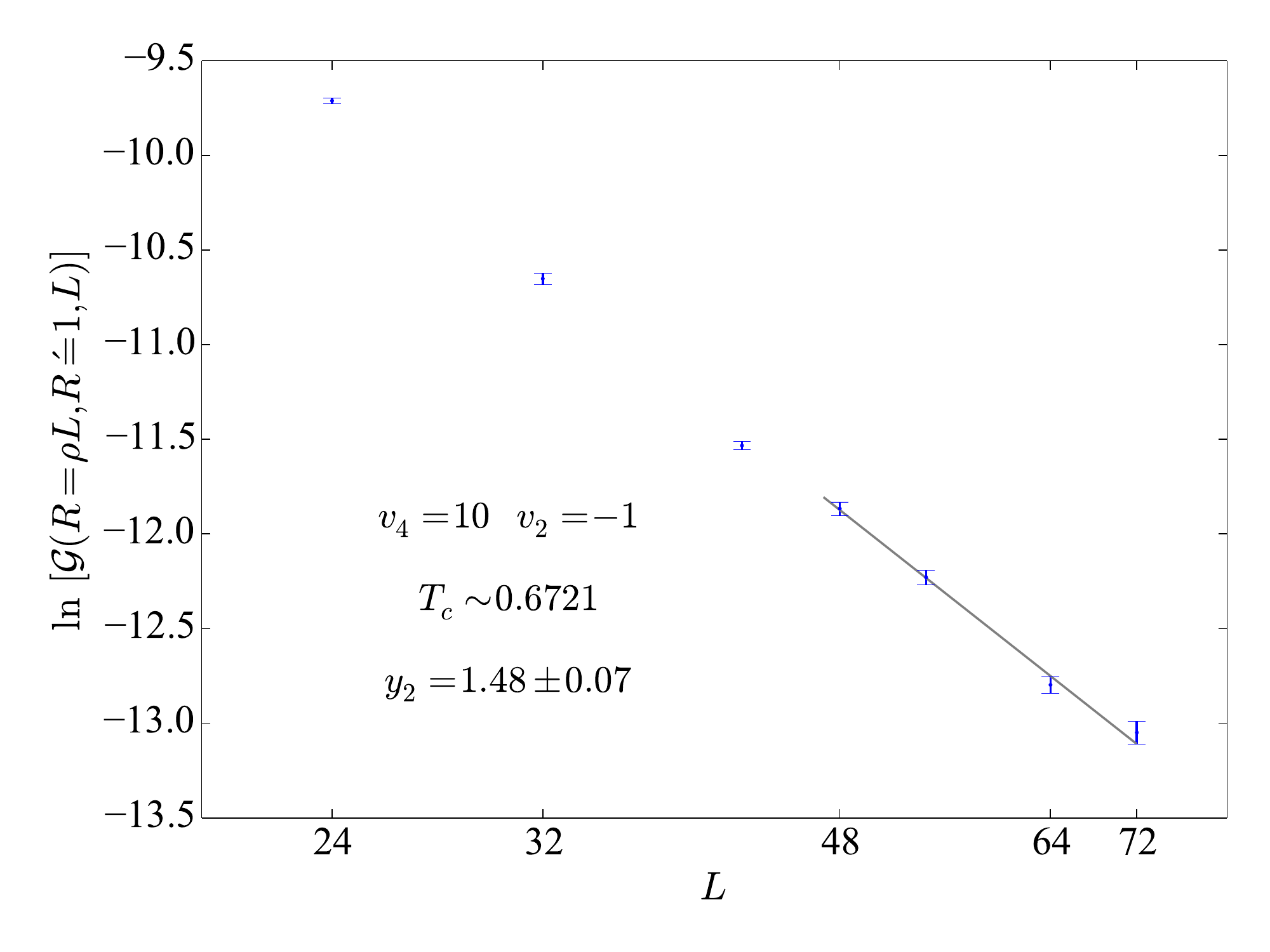}
\caption{Log--log plot of $\scG_2(R=\rho R_{\rm max},R'=1,L)$ for various system sizes at the critical temperature, with $v_4 = 1.0$ (top), $1.5$ (middle), and $10$ (bottom) and $\rho = 0.95$. The RG eigenvalue $y_2$ is related to the slopes \(s\) of the lines as $s=-2x_q=-2(d-y_q)$.}
\label{scaling_charge_2}
\end{figure}

\reffig{Q2pd} shows the pair correlation function \(\scG_q\) for charge $q=2$, rescaled by multiplying by $L^{2x_2}$ and plotted against ${R}/{L}$. The scaling dimension $x_2$ can be estimated from the plot in \reffig{scaling_charge_2}, showing $\scG_2(R=\rho R_{\rm max}, R'=1,L)$ as a function of $L$ for $\rho={R}/{L}=0.95$, using \refeq{EqscGscaling1}. The calculated RG eigenvalues are $y_2=1.58(2)$, $1.54(8)$, and $1.48(7)$ for $v_4=1.0$, $1.5$, and $10.0$ respectively, in agreement within error bars.

The dashed lines in \reffig{Q2pd} show a function \(\propto (R/L)^{2x_2}\), using the values of \(x_2\) calculated in \reffig{scaling_charge_2}. Using \refeq{EqScalingFixedL}, one expects the slopes of the lines to match those of the scaled distribution function in a region with \(1 \ll R \ll L\). We find slightly larger values for \(x_2\) by this method, and an estimate of \(y_2 \simeq 1.4\) obtained by analyzing the scaling with $b$ around $R=0.11 L$. The narrow region in which the scaling form applies, particularly for the relatively small system sizes that are accessible here, precludes a more precise estimate using this method.

\subsection{Charge 3}

\begin{figure}
\includegraphics[width=\columnwidth]{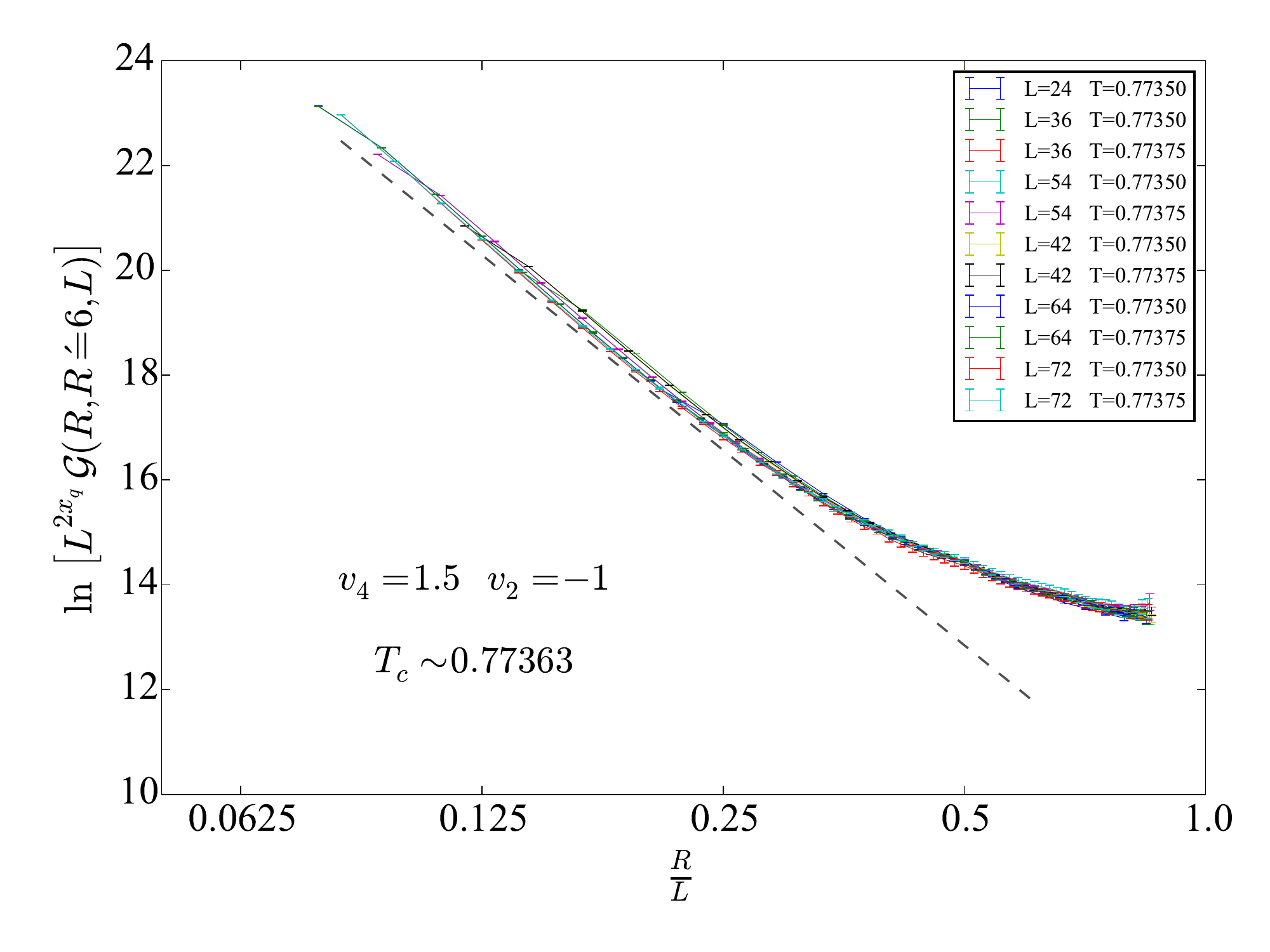}
\includegraphics[width=\columnwidth]{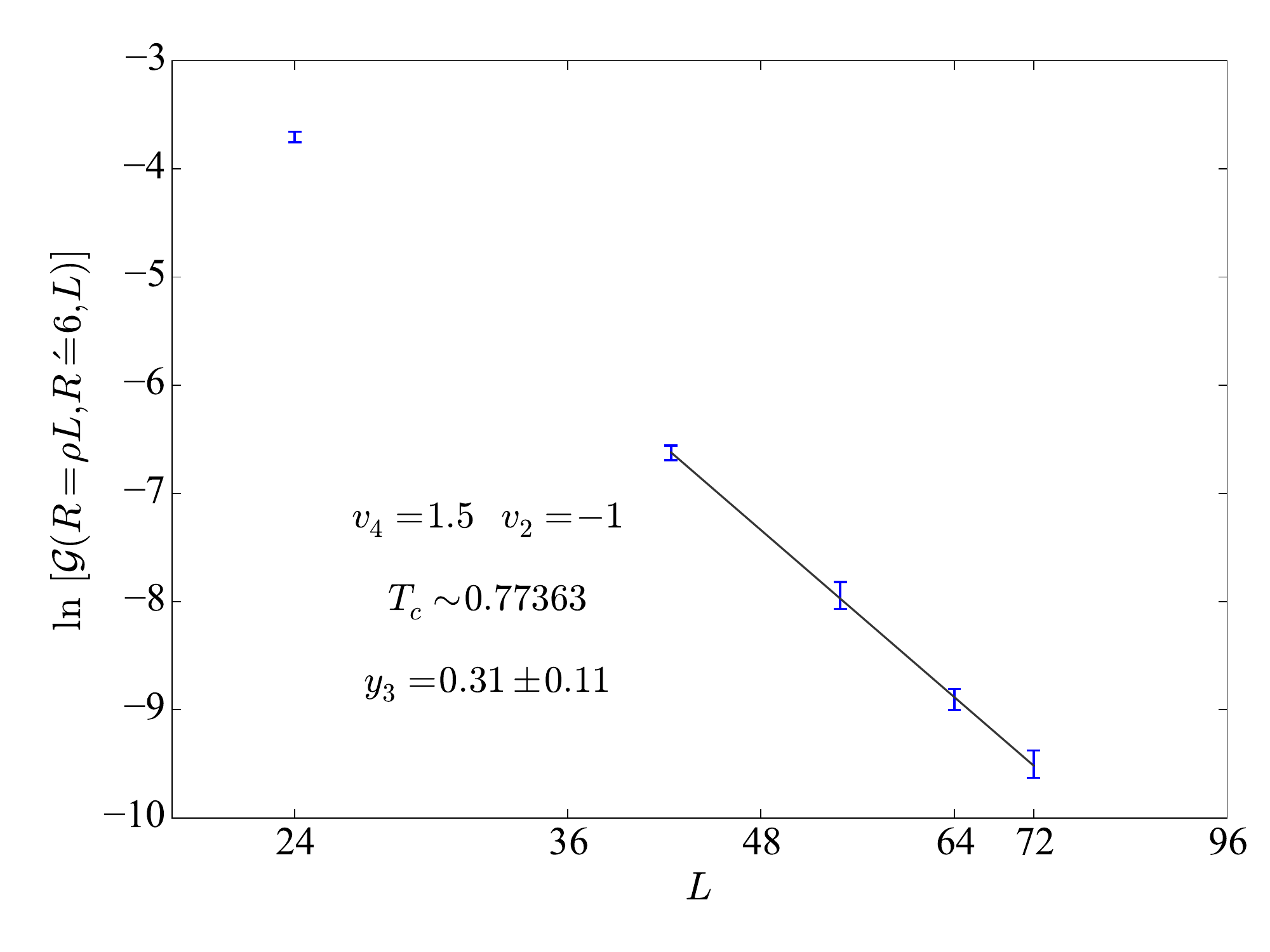}
\caption{Top: Pair correlation function for charge $q=3$, using piecewise estimation. As for the case of charge $2$ in \reffig{Q2pd}, the function has been rescaled to show the scaling with $L$. The dashed line has a slope of $-2x_3$. Bottom: Log--log plot of $\scG_2(R=\rho L,R'=6,L)$ versus $L$. The scaling dimension can be inferred directly from the slope of the best-fit line.}\label{scaling_charge_3}
\end{figure}
The scaling dimension of the $q=3$ charge was obtained using both of the methods described in \refsec{SecImprovingConvergence}. Results using piecewise estimation, presented in \reffig{scaling_charge_3}, give an RG eigenvalue of $y_3=0.31(11)$. The calculations were performed for $v_4=1.5$ and the scaling dimension was obtained using the monopole distribution function at a distance of $R=0.95R\sub{max}$. In this case, reasonable agreement is found with \refeq{EqScalingFixedL} for \(5 \lesssim R \lesssim 0.25L\).

\begin{figure}
\includegraphics[width=\columnwidth]{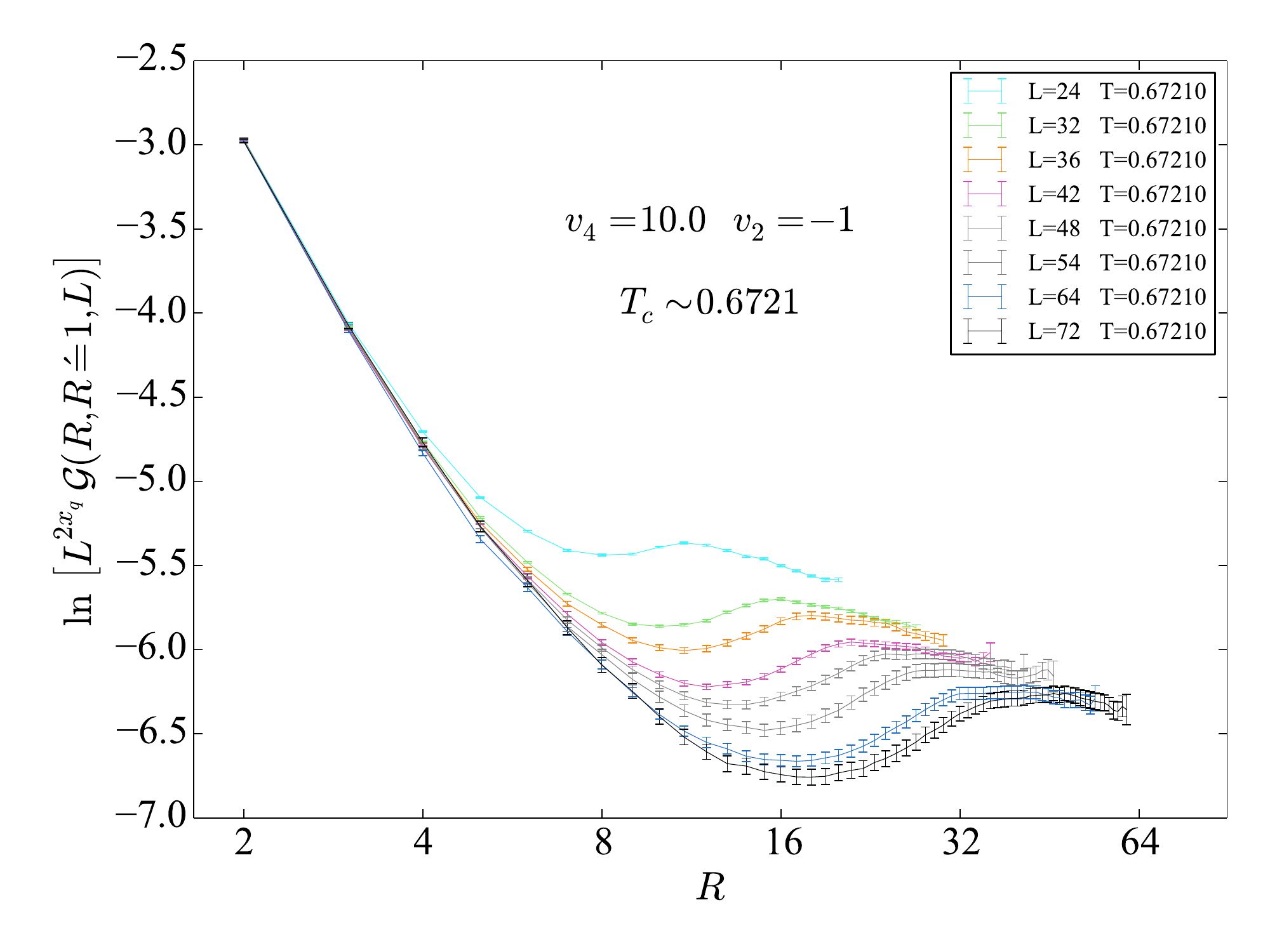}
\caption{Pair correlation function $\scG_3(R=\rho L,R'=1,L)$ calculated with an additional repulsive interaction between the charges of the form given in \refeq{EqPhiL} with $\theta=\frac{5}{2}$.}
\label{pcFAddedRep}
\end{figure}
\begin{figure*}
\includegraphics[width=\columnwidth]{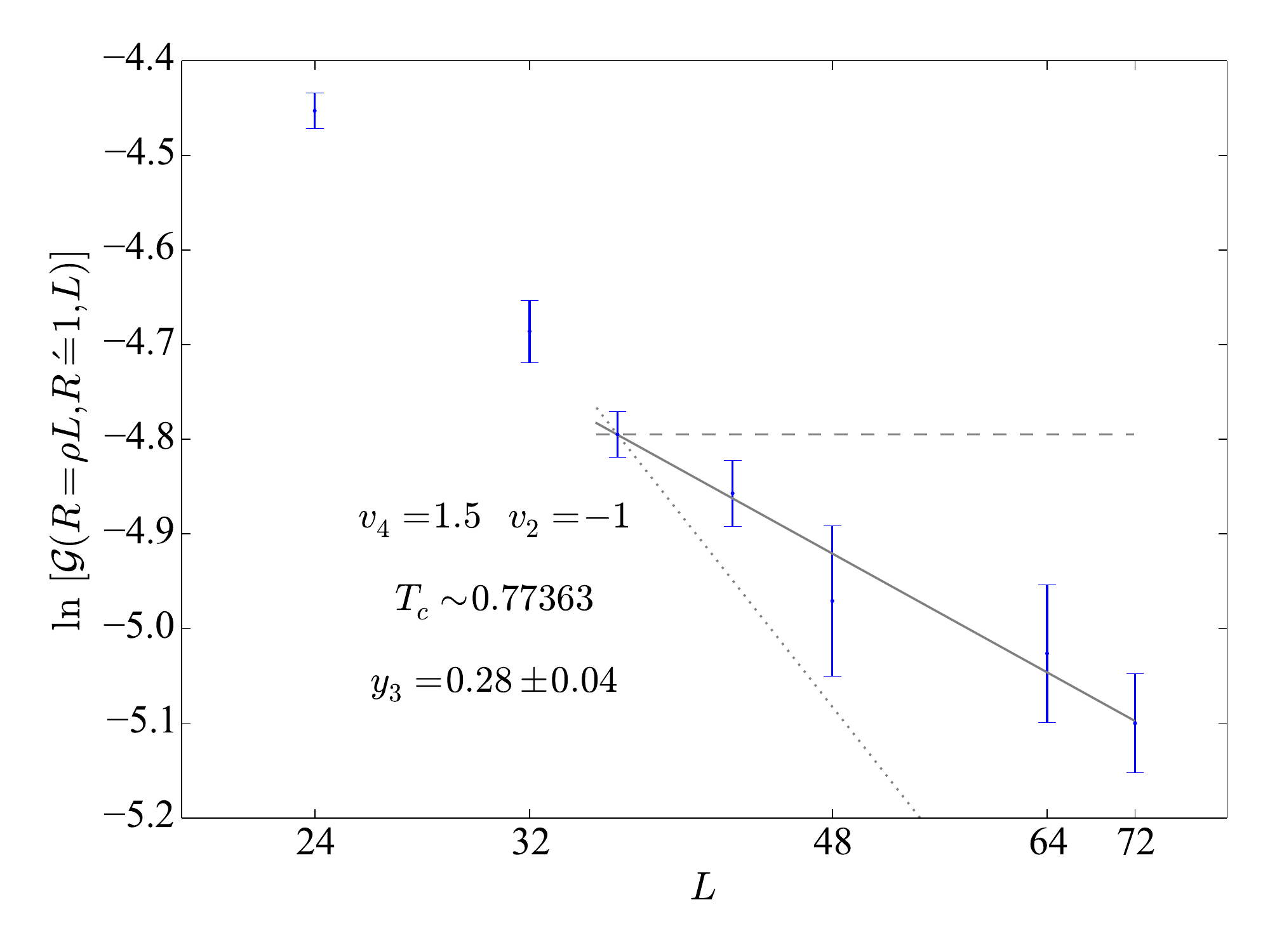}
\includegraphics[width=\columnwidth]{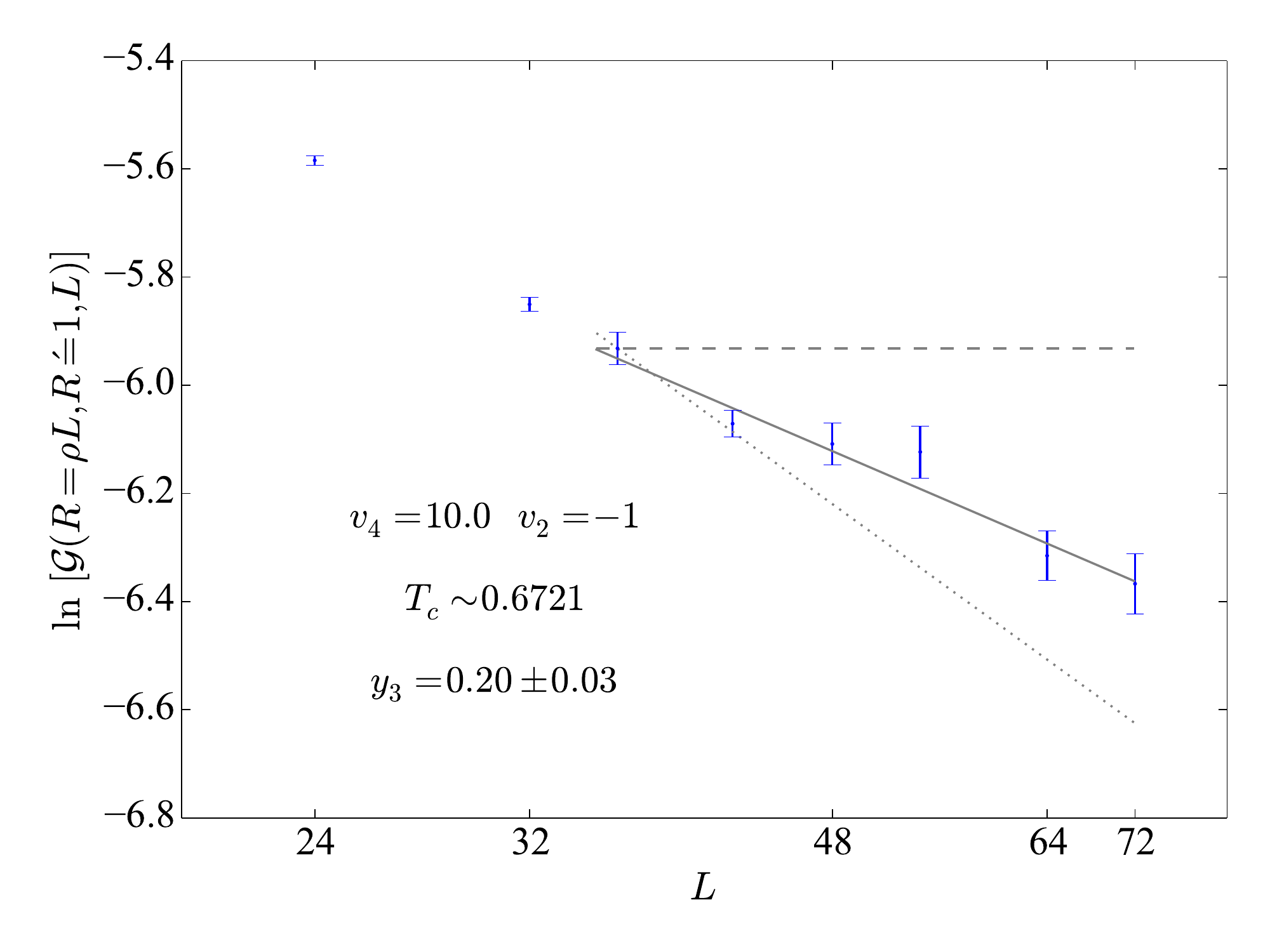}
\caption{Log--log plot of $\mathcal{G}_3(R=\rho L,R'=1,L)$ versus $L$, for \(v_4 = 1.5\) (left) and \(10.0\) (right) and $\rho = 0.95$, calculated with an added repulsive potential. The continuous lines show an error-weighted least-square fit. The dotted and dashed lines correspond to the cases where $y_3$ would have been $0$ and $0.5$ respectively. The RG eigenvalue \(y_3\) is related to the slope \(s\) by $s=-2(d-\theta-y_3)$, where \(\theta = \frac{5}{2}\) parametrizes the additional interaction.}
\label{scaling_charge_3_addedRep}
\end{figure*}

In order to obtain a better estimate of the exponent, we performed the calculations with an added repulsive interaction for $v_4=1.5$ and $v_4=10$. \reffig{pcFAddedRep} shows the plot of $\scG_3$ for this case. With the added repulsive interaction, short distance features appear to be amplified but the tail of the function still scales with the an exponent close to our previous estimate. The RG eigenvalues obtained were $y_3 = 0.28(4)$ and $0.20(3)$ for $v_4=1.5$ and $10$ respectively, as shown in \reffig{scaling_charge_3_addedRep}.

\section{Conclusions}
\label{SecConclusions}

We have demonstrated a method for calculating scaling dimensions of monomers with effective charge \(q\) in a classical dimer model. We have applied it to the columnar ordering transition of dimers on the cubic lattice and calculated the RG scaling eigenvalues \(y_q\) for monomers of charge \(q = 2\) and \(3\). For \(q = 2\), we find the values \(y_2 = 1.58(2)\), \(1.54(8)\), and \(1.48(7)\) for \(v_4 = 1.0\), \(1.5\), and \(10.0\) respectively. For \(q = 3\), we find \(y_3 = 0.28(4)\) and \(0.20(3)\) for \(v_4 = 1.5\) and \(10.0\) respectively. For comparison, the scaling eigenvalue for monomers of charge \(q = 1\) was found as \(y_1 = 2.421(8)\) for \(v_4 = 1.0\) in \refcite{Sreejith} (where it is denoted \(y_z\)), using the standard directed-loop algorithm.

The slight drifts in the scaling dimension with \(v_4\) are consistent with previous results for this transition \cite{Charrier2}, and have been attributed to corrections to scaling resulting from a nearby tricritical point \cite{CubicDimersPRB}. The values for the largest \(v_4\), furthest from the tricritical point, should therefore be considered most reliable. Recent work \cite{Nahum2} has indicated that this universality class exhibits unusual finite-size effects, providing an alternative explanation for this behavior. As noted in \refsec{SecResultsCharge2}, we indeed observe small discrepancies between results obtained with and without assuming standard finite-size scaling, but considerably larger system sizes are required to clarify this point.

There is now quite strong theoretical and numerical support for the claim that this transition is in the \NCP\ universality class, and so other transitions in the same class should have identical exponents. Of most current interest are quantum phase transitions in 2D \(S=\frac{1}{2}\) antiferromagnets \cite{Senthil}, such as the \JQ\ model \cite{SandvikJQ}, between N\'eel and VBS phases. The scaling dimensions of monopoles of charge \(q\), and, in particular, whether they are relevant or irrelevant, are of crucial importance for determining the fate of such transitions. For the rectangular (i.e., with twofold rotational symmetry) and square lattices, the situation is relatively clear: The nature of the transition in the rectangular-lattice \JQ\ model is determined by the sign of \(y_2\). The latter is clearly positive, and so the transition is certainly not described by the \NCP\ class. For the square-lattice \JQ\ model, the nature of the transition depends on the sign of \(y_4\). Previous results of quantum MC simulations on this model \cite{Lou} generally suggest that the transition is indeed continuous.

For this reason, we have concentrated here on \(y_3\), which is applicable to the honeycomb lattice, where the sites have threefold rotation symmetry and the smallest allowed monopole charge is \(q\sub{min} = 3\). The positive value of \(y_3\) indicates that the N\'eel--VBS transition on the honeycomb lattice should not be in this universality class, and would most likely be driven first order. Quantum MC results for such systems \cite{Pujari,Harada,Pujari2} have seen no clear evidence of a first-order transition, although a trend in this direction with increasing system size has been suggested \cite{Harada}. Our results indicate that the true nature of this transition is indeed first order, but the small value of \(y_3\) is consistent with critical behavior, described by the \NCP\ universality class, over a range of length scales.

Even allowing for the large and unconventional finite-size effects in this system \cite{Charrier2,Nahum2}, it seems unlikely that \(y_3\) would be more than very weakly irrelevant, which would mean that three-fold anisotropy should be visible over moderate length scales in the \JQ\ model on the honeycomb lattice. Pujari et al.\ \cite{Pujari2} have recently reported evidence of such anisotropy at the critical point.

While the observed order of the transition in the \JQ\ model provides evidence for the sign of \(y_q\), few quantitative values of \(y_q\) for \(q > 1\) are available with which to compare. By studying the scaling of powers of the VBS order parameter directly in the quantum model, Harada et al.\ \cite{Harada} found \(y_2 \simeq 1.0\). One can also compare with large-\(N\) expansions of the \(CP^{N-1}\) model \cite{MurthySachdev,Dyer}, which give (in our notation) \(x_2 = 0.311N - 0.234 + \bigO(N)\) and \(x_3 = 0.544N + \bigO(N^0)\). Truncating the expansions at these orders and setting \(N = 2\) gives \(y_2 = 2.612\) and \(y_3 = 1.912\), but there is of course no reason to expect the truncation to be reasonable in this case.

It would also be of interest to obtain an estimate on similar lines for the monopole with \(q=4\), which is of relevance for the \JQ\ model on the square lattice. In this case, however, single-site defects, analogous to those in \reffig{ChargesIllustration23}, are even more difficult to move around, severely limiting the accessible system sizes. One can, in principle, construct an alternative algorithm where each charge 4 is realized as a fusion of two charge-$2$ monopoles situated on the nearest sites of same sublattice. Our preliminary results for this case show significant direction dependence in even the largest systems we studied ($L=72$), making it difficult to obtain any useful estimates. Nonetheless, the systematic decrease of \(y_q\) with increasing \(q\) (for \(1 \le q \le 3\)) and the small value of \(y_3\) suggest that \(y_4\) is likely negative, consistent with observations of a continuous transition of the square-lattice \JQ\ model \cite{Lou}.

\acknowledgments

We are grateful to Ribhu Kaul and Anders Sandvik for helpful discussions, and to NORDITA for hosting the program ``Novel Directions in Frustrated and Critical Magnetism'' (July--August 2014), where some of this work was carried out. The simulations used resources provided by the Swedish National Infrastructure for Computing (SNIC) at the National Supercomputing Centre (NSC) and High Performance Computing Center North (HPC2N), and by the University of Nottingham High-Performance Computing Service.

\end{document}